\documentclass{elsarticle}

\usepackage[latin1]{inputenc}
\usepackage{amsmath,amsfonts,amssymb}
\usepackage{a4,theorem}
\usepackage{latexsym}
\usepackage{epsfig}

\usepackage{graphicx}
\usepackage{epstopdf}
\newtheorem{prop}{Proposition}[section]
\newtheorem{definition}[prop]{Definition}
\newtheorem{remark}[prop]{Remark}
\newtheorem{proposition}[prop]{Problem}
\newtheorem{theorem}[prop]{Theorem}
\newtheorem{lemma}[prop]{Lemma}
\newtheorem{example}[prop]{Example}
\newtheorem{corollary}[prop]{Corollary}

\begin{document}

\begin{frontmatter}

\title{Plasmid segregation and accumulation}

\author[TUMGarching,ICB]{Johannes M\"uller\corref{mycorrespondingauthor}}
\author[TUBS]{Karin M\"unch}
\author[TUMGarching]{Bendix Koopmann}
\author[TUMGarching]{Eva Stadler}
\author[TUBS]{Louisa Roselius}
\author[TUBS]{Dieter Jahn}
\author[TUBS]{Richard M\"unch}

\address[TUMGarching]{Center for Mathematics, Technische Universit\"at M\"unchen, 85748 Garching, Germany}
\address[ICB]{Institute for Computational Biology, Helmholtz Center Munich, 85764 Neuherberg, Germany}
\address[TUBS]{Institute of Microbiology and Braunschweig Integrated Centre of Systems Biology (BRICS), Technische Universit\"at Braunschweig,  D-38106 Braunschweig, Germany}

\cortext[mycorrespondingauthor]{Corresponding author}

 \begin{abstract}
The segregation of plasmids in a bacterial population is investigated.  
Hereby, a dynamical model is formulated in terms of a size-structured population
using a hyperbolic partial differential equation incorporating non-local terms (the fragmentation equation). 
For a large class of parameter functions this PDE can be re-written as an infinite system of ordinary differential equations for the moments of its solution.
We investigate the influence of different plasmid production modes, kinetic parameters, and plasmid segregation modes 
on the equilibrium plasmid distribution. 
In particular, at small plasmid numbers the distribution is strongly influenced by the production mode, while the kinetic parameters (cell growth rate resp.\ basic plasmid reproduction rate) influence the distribution 
mainly at large plasmid numbers. 
The plasmid transmission characteristics only gradually influence the distribution, but may become of importance for biologically relevant cases. 
We compare the theoretical findings with experimental results.
 \end{abstract}
 

\end{frontmatter}

\newcommand{\field}[1]{\mathbb{#1}}
\newcommand {\R}{\field{R} }\newcommand {\N}{ {\field{N}} }
\newcommand {\smallN}{\field{N} }\newcommand {\C} {\field{C} }
\newcommand {\Q}{\field{Q} }\newcommand {\Z}{\field{Z} }
\newcommand {\FF}{{\cal F} }\newcommand {\Oo}{{\cal O} }
\newcommand {\Ss} {{\mathfrak S}}\newcommand {\ess} {\mbox{{\rm ess}} }
\newcommand {\supess} {\mathop{\mbox{{\rm supess}}} }
\newcommand {\supp}{{\rm supp} }

\newcommand {\kommentar}[1]      {}


 
\section{Introduction}

Plasmids are self-replicating, extra-chromosomal DNA molecules most commonly found in bacteria.
Genes coded on naturally occurring plasmids typically support the survival under various environmental conditions such as antibiotic resistance, specific degradation pathways, virulence, amongst others.
In genetics and biotechnology plasmids serve as important tool to express particular genes e.g. for the recombinant production of proteins~\cite{Terpe2006}.  
In low-copy plasmids the copy number ranges from 1-2 copies per cell.
During cell division those plasmids are actively segregated like chromosomes via a so called partitioning system. 
The copy number of high-copy plasmids can be up to several hundred molecules per cell.
Although it is questioned, the general assumption is, that high-copy plasmids without partitioning system segregate stochastically by random diffusion~\cite{Million-Weaver2014}.

The dynamics of the plasmid distribution in a bacterial  population is 
of interest, e.g., in order to understand the spread of new properties, but also in order to optimize processes in biotechnological engineering. 
In particular, recent experimental findings show an accumulation of high copy plasmids
in some cells~\cite{manuskriptKarin}. 
To understand the background of this accumulation is 
of large interest, as for biotechnological production techniques, neither cells with only few plasmids nor cells with too many plasmids are desired: 
cells with only few plasmids do not produce efficiently as they are insufficiently
triggered to do so, and cells with too many plasmids will not 
produce well as the metabolic costs for plasmid reproduction are too high.\par

Modeling plasmid distributions has a rather long tradition. 
Most approaches are based on simulation models of a population structured by the
number of plasmids per cell~\cite{bentley1993,kuo1996,goss1999,nordstroem2006}. 
Only few articles go to a continuum limit, and consider a hyperbolic partial differential equation~\cite{gusanov2000}. 
All these models indicate that an unimodal
distribution should be expected, resembling a gamma distribution. 
These papers do not address plasmid accumulation, but rather the condition 
of plasmid loss.
\par
Experimental findings show that high copy plasmids are in general not equally distributed among the daughter cells, but often one of the
daughters receives systematically 
a higher fraction of plasmids than the 
other daughter~\cite{manuskriptKarin}.
 The implication of the 
characteristics of such an unequal transition
from mother to daughter is unclear and
under discussion~\cite{kim1991}. Summers and Sheratt 
conjecture 
 that this unequal transition also influences the 
plasmid distribution in a crucial 
way~\cite{summers1984}. 
One  hypothesis states that this
characteristics is a key mechanism that 
leads to aggregation of many plasmids in 
some cells.
\par
In order to address this hypothesis, we propose a simple model that 
focuses on the basic mechanisms. 
We do not consider horizontal transmission 
of plasmids (neither directly from cell to cell
by competent cells, nor by {\it de novo} infection 
due to environmental plasmids), 
but only vertical spread from mothers to daughters. 
We also do not assume that the plasmid load affects the population dynamics of bacteria. We focus in particular on the interplay between plasmid and 
bacterial reproduction, 
taking into account the characteristics
of plasmid segregation. 
 We treat plasmids as 
an infectious disease that is exclusively spread by vertical 
transmission, that is, the infectivity is taken to zero. In contrast to epidemic
models (and plasmids can be considered as infectious agents), in our model 
plasmid spread and cell reproduction are strongly intertwined, which leads to
difficulties in separating information about population and plasmid dynamics. We obtain a hyperbolic 
partial differential equation, the so-called fragmentation
equation,  
with a structure close cell size models~\cite{diekmann:cellSize,gyllenberg1987,wake2014,zaidi2015}, 
see also the book of Perthame~\cite{perthame:book} and quotations therein. A lot
of work is done for cell size models, in particular the
asymptotic behaviour is well known. We give some overview about 
the most important existence results in section~\ref{exResOver}. 
In the present paper, the primarily aim is not to extend 
results about existence and asymptotic stability of 
an equilibrium 
distribution, but 
 aim at a characterization of its asymptotic shape.\\
In order to disentangle plasmid and cell population dynamics,  we  focus on moments. The zero'th moment corresponds to the population size, the first
moment to the amount of plasmids within the population etc. We find an infinite systems of ordinary differential equations 
for these moments. \\
At this point, we introduce two fundamentally different production mechanisms for plasmid  production: In 
the plasmid-controlled
mode (also called mass-controlled mode~\cite{kuo1996}) 
each plasmid reproduces itself, such that the reproduction 
rate is in lowest order proportional to the plasmid number. In the cell controlled mode, basically the cell determines 
 the plasmid production rate, such that 
the rate is to a large extend independent on the number of plasmids present. 
For the plasmid controlled mode, 
we 
obtain a fundamental threshold theorem: 
the population loses plasmids if 
cell reproduction is faster than plasmid reproduction. 
Threshold theorems of this type are well known from the theory of communicable diseases. In the cell controlled
mode, plasmids are of course never lost. 
Next we concentrate on the equilibrium distribution of plasmids in the population. 
In particular we aim to identify reasons that lead to aggregation of 
plasmids in cells. It turns out that aggregation 
of plasmids is first of all influenced by the 
ratio between basic plasmid reproduction rate and cell reproduction rate. If this factor is less 
than two, we find 
that the distribution tends to zero at the 
carrying capacity of plasmids. If this factor exceeds two, a singularity builds up at
the carrying capacity: the distribution tends to infinity, indicating aggregation of many plasmids in some cells. The characteristics of plasmid
segregation does influence this shape, but is only of  minor importance.\par
The paper is structured as follows: 
in section~\ref{modelSec} we introduce the 
discrete model and the continuum limes that yields the hyperbolic
partial differential equation and 
discuss in section~\ref{exResOver} some relevant literature and state
some simple properties of the equation.
 In section~\ref{theo} we 
reformulate the PDE in terms of moments, and analyze the shape
of plasmid distribution in 
several scenarios for the plasmid reproduction. 
We deepen this discussion of the influence 
of parameters 
on the plasmid distribution in section~\ref{shapeSect},
basically by means of numerical simulations. Section~\ref{experi} 
compares theoretical and experimental results, and in the last section~\ref{discuss} we discuss our findings.

\section{Model} \label{modelSec}
\subsection{Discrete Model}

\begin{figure}[tbh]
\begin{center}
\includegraphics[width=10cm]{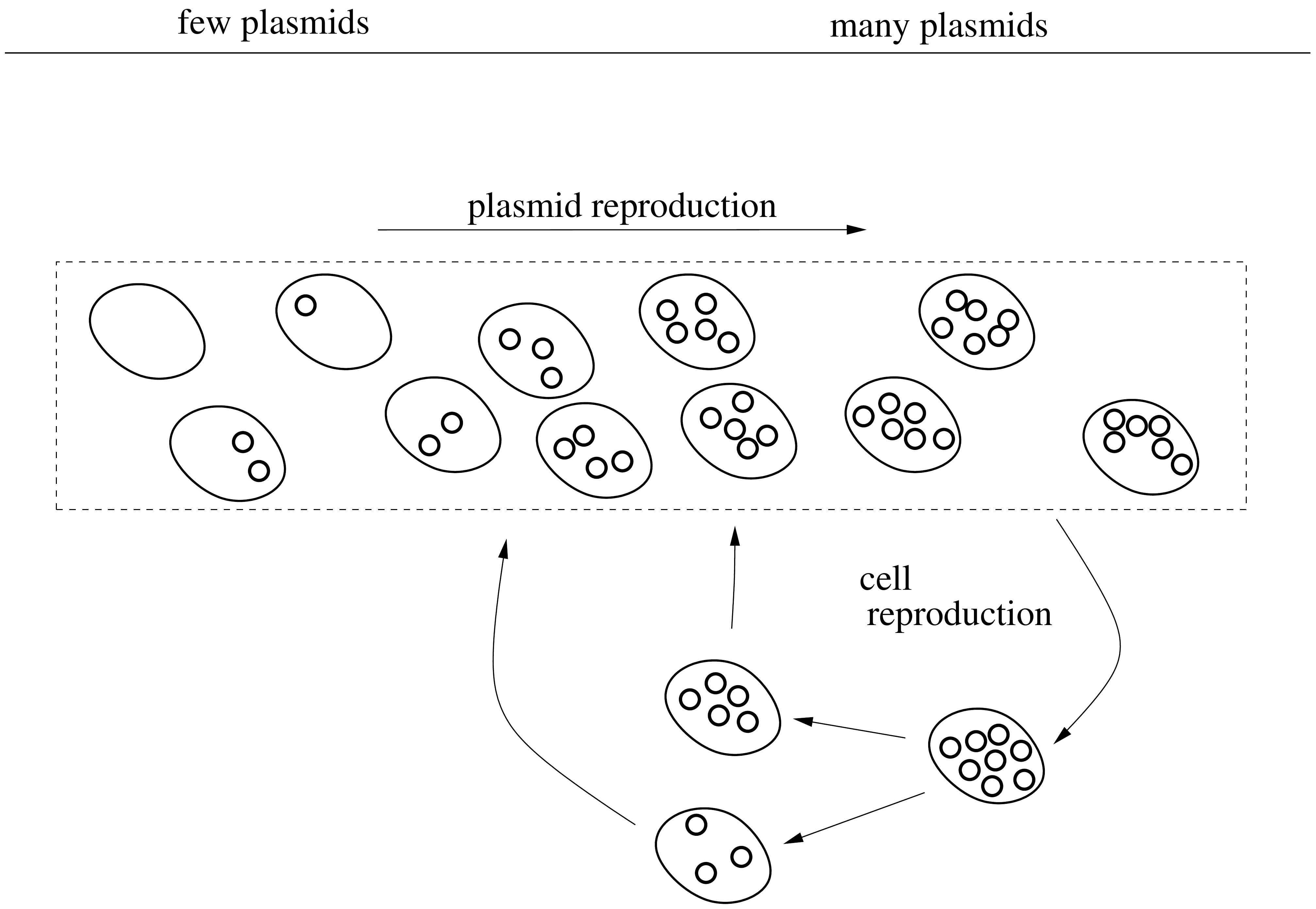}
\end{center}
\caption{Mechanisms implemented in the model: plasmids reproduce 
within cells, cells divide, and during cell division plasmids are transferred from mother to daughters.}\label{model}
\end{figure}

We start with a model discrete in state, similarly to~\cite{bentley1993}.
The population size of bacteria containing $i$ plasmids 
at time $t$ 
is denoted by $x_i(t)$ (see also table~\ref{paraExplain} for the meaning of the parameters).
The processes that mainly affect the dynamics of $x_i$ are 
cell division (and cell death) that decrease the number of 
plasmids per cell, and plasmid reproduction 
that increases the plasmid number per cell. 
Cell- and plasmid reproduction counteract 
and their interplay determines the plasmid distribution 
(see figure~\ref{model}). 
The reproduction 
rate of plasmids within a cell already 
containing $i$ plasmids is $\tilde b(i)$. Later, we will specify 
$\tilde b(i)$ such that different scenarios can be analyzed. \par

\noindent
Cells die at rate $\mu$, and reproduce at rate $\beta$. 
We assume that neither the death- nor 
the division rate is affected by the number of plasmids contained by a cell. 
At this point, it would be simple (and more realistic) to include a  plasmid-dependence of these rates.  
However, as we restrict ourselves exclusively to constant rates
 in the present work, we stay with this basic case.\\   
Next the transmission of 
plasmids from mother to daughter is modeled. Very often, bacteria
are shaped like small cylinders. One side of the 
cylinder (bottom or top) is tagged - this is the pole 
of the cell. One of the daughter cells inherits the 
old pole of the mother cell. In this way, the two daughter cells
can be distinguished. If the mother contained $i$ plasmids, 
we denote by $p(j;i)$ the probability that the daughter 
inheriting the mother's pole obtains $j$ plasmids 
(consequently, the other daughter receives $i-j$ plasmids). 
The average number of
daughters containing $j$ plasmids in case that the 
mother contains $i$ plasmids is given by 
$p(j;i)+ p(i-j;i)$. We will identify different 
transmission characteristics and investigate their effect. 
Since we do not distinguish between the daughter cell incorporating the mothers pole and that incorporating a new pole, even an unequal segregation leads to 
a symmetric shape of $p(j;i)+ p(i-j;i)$. Unequal segregation is likely to produce a bimodal shape, while equal segregation will in most biologically relevant cases result in an unimodal shape. \\ 
Cells divide at rate $\beta$, and in this, leave their population class
$x_i$. This process yields a term $-\beta x_i$ in the model. The daughter
cells are distributed to population classes with less or equal many plasmids. 
Equivalently, all daughters that receive $i$ plasmids enter $x_i$. 
Their mothers necessarily contained $i$ or more plasmids. 
These model assumptions lead 
to a description of the cell division 
(and only cell division) by the equation
$$ x_i' =  -\beta x_i + \beta \sum_{i'=i}^\infty [p(i;i')+p(i'-i;i')] x_{i'}.$$
We formally extend the sum to infinity, though in biologically relevant situations 
we expect a maximal number of plasmids a cell 
will contain, such that eventually $x_i=0$. 
Note that 
\begin{eqnarray}
  \sum_{i=0}^{i'} [p(i;i')+p(i'-i;i')] =2\label{pobSum}
\end{eqnarray}
and 
$$ \sum_{i=0}^{i'} i\,[p(i;i')+p(i'-i;i')] 
 =   \sum_{i=0}^{i'} i\,p(i;i')-\sum_{i=0}^{i'} (i'-i)p(i'-i;i') +  \sum_{i=0}^{i'} i'\,p(i;i')\nonumber\\
  = i'.
$$
These two equations indicate that the number of cells is doubled 
and the number of plasmids is conserved in cell divisions.\\
All in all, we obtain for $x_i$ the system of ordinary 
differential equations
\begin{eqnarray}
x_i' &=& -(\beta+\mu) x_i 
+  \tilde b(i-1)\, x_{i-1} - \tilde b(i)\, x_i\label{discreteModelEq}
\\
&&\qquad\qquad\qquad\qquad+ 
\beta \sum_{i'=i}^\infty [p(i;i')+p(i'-i;i')] x_{i'}\nonumber
\end{eqnarray}
where we formally take $x_{-1}(t)\equiv 0$. 
As the experiments we 
will consider  below are performed for high copy plasmids 
(that is, $x_i\not = 0$ for $i$ large),
we proceed towards  the continuum limit. 
Since also the cell population is large, stochastic effects can be assumed to be negligible.

\begin{table}[ttb]
\begin{center}
\begin{tabular}{l|l|l}
    & discrete model & continuous model\\
    \hline
 Amount of plasmids & $i$ & $z$ \\
 Population size at time $t$ & $x_i(t)$ & $u(z,t)$\\
 Cell division rate & $\beta$ & $\beta $\\
 Cell death rate & $\mu$ & $\mu $\\
 Plasmid reproduction rate & $\tilde b(i)$ & $b(z)$\\
 Plasmid transmission kernel & $p(i; i')+p(i'-i; i')$ & $k(x,y)$
\end{tabular}
\end{center}
\caption{Parameters for the discrete and the continuous model.}
\label{paraExplain}
\end{table}

\subsection{Continuum limit}
In order to proceed to the continuum limit, 
we assume at the time being that there is a smooth function 
$u(z,t)$ such that for $h$ small
$$x_i(t) \approx \int_{i\,h-h/2}^{i\, h+h/2} u(z,t)\, dz \approx u(i\, h,t)\, h, 
$$
and that
 $\tilde b(i) \approx b(h\,i)/h$, 
$p(i;i')+p(i'-i;i') \approx h\,k(i\,h,i'\, h)$. 
Eqn.~(\ref{pobSum}) indicates that
\begin{eqnarray}
 \int_0^y k(x,y)\, dx = 2\label{k2integral}
 \end{eqnarray}
and its definition implies a certain symmetry,
$$k(y-x,y) = k(x,y).$$
In the limit $h\rightarrow 0$, we obtain
\begin{eqnarray}
\partial_t u(z,t) + \partial_z(b(z) u(z,t)) &=& -(\beta+\mu) u(z,t) \label{mod0X3} 
+ \beta \int_{z}^\infty k(z,z') u(z',t)\, dz'
\end{eqnarray}
with  $u(z,t)\in L^1$ for $t$ given. 
The plasmid reproduction rate $b(z)$ is the flux of this hyperbolic partial differential equation. 
We superimpose zero flux boundary conditions 
at $z=0$, 
\begin{eqnarray}
 b(0) u(0,t) = 0.   
 \end{eqnarray}
If $b(0)=0$, this condition is trivial, but we
also intend to consider the rather extreme case that
$b(z)\equiv b_0$ is constant. If $b(z)$ posses a strictly positive limit for $z\rightarrow0+$, this boundary condition is required. We obtain the fragmentation equation as stated e.g.\ 
in the book of Perthame~\cite{perthame:book}.
\par\medskip

\begin{remark} 
The structure of the model is unexpectedly different from 
most epidemic models, where the dynamics of the 
endemics is formulated as the fate of one subpopulation, 
the class of infected individuals. 
Here, instead, the dynamics (and the parameters) of the infection are hidden in the flux of the size structure. 
Epidemic models close to the present approach address e.g.\ explicitly 
the parasite load per individual~\cite{hadeler1983}. Nevertheless, there is 
still a fundamental difference in spreading parasites and plasmids: 
An individual can transmit parasites horizontally to other individuals, while in our restricted model plasmids can be only passed from mother to daughter (vertical transmission). Standard epidemic models incorporate 
some aspect of irreducibility missing in the present context. 
In that aspect, a model that allows for plasmid release and uptake may
be even more simple to handle than the present one.\\ 
One central issue below will
be to separate and compare information about the population dynamics of bacteria and population dynamics of plasmids.
\end{remark}

\section{Existence results, simple properties}
\label{exResOver}
Fragmentation-aggregation processes did attract attention, particularly in recent years. We briefly indicate known results, particularly existence results; the main focus of the present paper is not existence results but properties of the long term behavior. 
\par\bigskip
The fragmentation-aggregation 
equation we obtained 
has a structure close to cell size models~\cite{diekmann:cellSize,gyllenberg1987,wake2014,zaidi2015}, 
see also the book of Perthame~\cite{perthame:book} and quotations therein. A lot
of work is done for cell size models, in particular the
asymptotic behavior is well known. 
In cell size models, a singularity appears if cells reach 
a critical size, as the rate at which cells divide 
tends to infinity at this size. This is a singularity in the reaction term 
of the equation. In the present case, 
the singularity appears in the plasmid dynamics within
cells. That is, the flow of the hyperbolic equation becomes
singular (i.e., the flow becomes zero). As stated in~\cite{doumic2010}, often 
there is no biological justification to assume a non-singular flow in the transport 
equation, but only few articles address this
problem, see e.g.~\cite{michel2006,doumic2007,doumic2010,Mischler2015,Campillo2016}.\\
As the model is linear, we expect in non-pathological cases that 
the solution 
approximates in the long run an exponentially growing solution. It is central to study 
the eigenvalue problem 
\begin{eqnarray*}
 \partial_z(b(z)U(z)) &=& -(\beta+\mu+\lambda) U(z) + \beta \int_{z}^\infty k(z,z') U(z')\, dz'\\
 b(0) U(0) &=& 0,\qquad\lambda\in\R,\qquad U\geq 0, \quad U(z)\not\equiv 0.
\end{eqnarray*}
Often, in addition it is assumed that $U$ is at least continuous; this condition 
rules out solutions that have a point mass at zero, 
and therefore it is often possible to 
obtain uniqueness results 
(even if $b(0)=0$).  
Convergence of the initial value problem towards $e^{\lambda t}\,c\, U(z)$ 
for $c\in\R$ suited are available in the case that $k(x,y) = 2\,\delta(x-y/2)$, $b=1$, $\mu=0$, and $\beta$ is continuous~\cite{Perthame2005},
and can often be concluded by the ``general relative entropy method''
(see~\cite{perthame:book} for more general $k$).\\ 
Particularly, existence and convergence results are known in case that $b(z)$ 
is constant~\cite{perthame:book},  
$b(z) = z^\alpha$~\cite{michel2006,Calvez2012}, 
or $b(z)$ has compact support (under the assumption that $\beta(0)=0$)~\cite{doumic2007}. In~\cite{doumic2010}, more general parameter functions $b(z)$ are allowed, but e.g.\ 
$\beta(z)/b(z)$ needs to be integrable at $z=0$.
To our knowledge, the existence of the solution $U(z)$ 
in case of logistic growth $b(z)=b_0z(1-z/z_0)$ 
and $\beta$ constant is not exactly handled. As mentioned before,
we also do not address the existence problem, but give an illustrative 
example for special parameters, where $U(z)$ can be explicitly determined.
 \par\medskip

In view of the biological question we aim to answer, we focus on properties of $U(z)$. 
Before we start with this investigation, we note some obvious properties of the model. 
 As usual, we denote by $\R_+=\{x\geq 0\}$, 
$\|\varphi\|_{C^1(\R_+)}=\sup_{x\in\R_+}(|\varphi(x)|+|\varphi'(x)|)$, and $L^1(\R_+)$ is the space of 
Lebesgue-integrable functions on $\R_+$. The first lemma indicates that 
an initial value with a compact support will always have a compact support.
  \par\medskip

\begin{lemma} \label{finiteSupport} 
{\bf (a)} Assume $b(z)=b_0z(1-z/z_0)$ or $b(z)=b_0(1-z/z_0)$,   
  $z_0>0$. Let $u_0(z)\in L^1(\R_+)$, 
$\supp(u_0)\subset [0, z_0-\varepsilon]$.  
Then, for all $t\in\R_0$ there is a smooth, 
monotonously decreasing function  $\varepsilon(t)>0$ such that 
 $\supp(u(z,t))\subset [0, z_0-\varepsilon(t)]$.\\  
 {\bf (b)} Assume $b(z)>0$ for $z>0$, and 
 $\|b(z)\|_{C^1(\R_+)}<\infty$.  Let 
$u_0(z)\in L^1$, where 
$\supp(u_0)$ is a compact interval.  
Then, for all $t\in\R_0$ 
$\supp(u_0)$ stays compact.   
\end{lemma}
{\bf Proof: } {\bf ad a.}
First of all, if $b(z_0)=0$ for some 
$z_0>0$ (first zero), then the position of the characteristic lines of the partial differential equation at hand indicates that 
$\supp(u(z,t)))\subset[0,z_0]$ for any initial
condition with support in $[0,z_0]$. Since we assume that 
$b(z)$ is differentiable, we even know that the support of 
any solution bounded away from $z_0$ may get arbitrary close but stays away from $z_0$ (in finite time). \\
{\bf ad b.} Since $\|b(z)\|_{C^1(\R)}<\infty$, the solution of the characteristic equations cannot blow up in finite time, 
and the support of $u(z,t)$ stays finite.
\qed\par\medskip

The simplicity of the population growth yields the following result.
\begin{proposition}\label{popDyn0}
Let  the assumption of lemma~\ref{popDyn0} be given.
Then, 
$$ \int_0^\infty u(z,t)\, dz = e^{(\beta-\mu)t} 
\int_0^\infty u(z,0)\, dz. $$
\end{proposition}
{\bf Proof: } 
Integrating equation~(\ref{mod0X3}) from zero to infinity  
is equivalent with integrating over a finite interval, since
the support of $u(z,t)$ for $t$ finite, given, is contained in 
a (growing, but for all times compact) interval. 
We find 
\begin{eqnarray*}
\frac d {dt}\int_0^\infty u(z,t)\, dz &= & 
-b(z)u(z,t)\big|_{z=0}^\infty
-
(\beta+\mu) \int_0^\infty u(z,t)\, dz \\
&&\qquad
+ \beta\int_0^\infty \int_z^\infty k(z,z')u(z')\, dz'\, dz\\
\end{eqnarray*}
With $b(z)u(z,t)\big|_{z=0}^\infty=0-b(0)u(0,t)=0$ and 
$$
\int_0^\infty \int_z^\infty k(z,z')u(z')\, dz'\, dz
=
\int_0^\infty \int_0^{z'} k(z,z')u(z')\, dz\, dz'
= 2 \int_0^\infty u(z,t)\, dz$$
we obtain 
\begin{eqnarray*}
\frac d {dt}\int_0^\infty u(z,t)\, dz &= & 
(\beta-\mu) \int_0^\infty u(z,t)\, dz 
\end{eqnarray*}
and the result follows.\qed
\par\medskip

Note that in general it is non-trivial to determine the 
exponential growth rate 
$\lambda$ (see e.g.~\cite{perthame:book}). The simplicity of our 
model assumptions yield the following proposition. 
\begin{corollary} Any solution of the form $u(z,t) = U(z)\, e^{\lambda t}$ with $U\in L^1(0,z_0)$ necessarily has exponent $\lambda=\beta-\mu$.
\end{corollary}

\section{Shape of the equilibrium plasmid distribution}
\label{theo}

We aim at the answer of two questions. (a) Are plasmids able to 
spread in the population, or does the 
average number of plasmids per cell tends to zero? 
(b)~If 
plasmids stay abundant, how does the stationary distribution of plasmids
look like? Respectively, can we identify factors that lead to an
accumulation of plasmids in some cells? \par\medskip

We will disentangle the dynamics of plasmids and cell population up to a certain degree in addressing not directly the solution $u(z,t)$ respectively
the function $U(z)$, but in focusing on the moments 
$M_i(t)=\int_0^\infty z^i\, u(z,t)\, dz$ respectively $P_i=\int_0^\infty z^i\, U(z)\, dz$. $M_0(t)$ indicates 
the total population size 
(regardless how many plasmids are present), 
$M_1(t)$ states the amount of plasmids contained by the total population (summarized over all cells), and $M_2(t)$ gives a hint 
about the variation of the plasmid distribution etc. In this way,
the different aspects (population dynamics of bacteria resp.\ plasmids) can be -- up to a certain degree -- considered separately.\par

In order to reduce 
the partial differential equation for $u(z,t)$ to an infinite
set of ordinary differential equations for $M_i(t)$, we first 
investigate moments of the kernel $k(x,y)$ 
(see also~\cite[Section 4.2]{perthame:book}
or  \cite[Section 1.2]{Mischler2015}  for the next 
 results).
\par\medskip

\begin{lemma}\label{k2m1}  Assume $k(x,y)=k(y-x,y)$, and 
 $\int_0^yk(x,y)\, dx = 2$ for $y>0$. Then, 
 $\int_0^yx\,k(x,y)\, dx =y$.
\end{lemma}
{\bf Proof: } 
We find
$$\int_0^yx\,k(x,y)\, dx
=
\int_0^y(y-x)\,k(y-x,y)\, dx
=
\int_0^y(y-x)\,k(x,y)\, dx
= 2 y - \int_0^yx\,k(x,y)\, dx.$$
Thus,  $\int_0^yx\,k(x,y)\, dx
=y$.\qed\par\medskip

\begin{definition} \label{scalDef} Let $k(x,y)\geq 0$, 
$\int_0^y|k(x,y)|\, dx<\infty$ for $y>0$. 
If the moments  of $k(x,y)$ satisfy
\begin{eqnarray}
 \int_0^y x^ik(x,y)\, dx = y^i \alpha_i
\end{eqnarray}
with $\alpha_i = \int_0^1\xi^i k(\xi,1)\, d\xi$, $\alpha_0=2$, 
$\alpha_1=1$, $\alpha_{i}>\alpha_{i-1}$, 
and $\sum_{i=1}^\infty \alpha_i/i<\infty$ for $i\rightarrow\infty$,
we call the kernel scalable.
\end{definition}

Note that it is straightforward to formulate this 
definition not only for integrable kernels but also for 
kernels consisting of distributions 
(e.g.\ $\delta$-peaks). The condition $\sum_{i=1}^\infty \alpha_i/i<\infty$ forces $\alpha_i$ to converge sufficiently fast to zero; we will use this fact later in the paper. 
The next lemma explains 
why we call kernels characterized by 
definition~\ref{scalDef} ``scalable''. In particular, we find 
that the condition $\sum_{i=1}^\infty \alpha_i/i<\infty$ corresponds
to a certain integrability condition for $k(x,y)$.

\begin{lemma} Let $k(x,y)>0$, and $\int_0^y\ln(1/(1-x))\,k(x,y)\, dx<\infty$ for $y\in[0,1]$. 
If  $\int_0^yk(x,y)\, dx=2$, $k(x,y)=k(y-x,y)$, and 
$$ k(x,y) = k(x/y, 1)/y,$$
the kernel $k(x,y)$ is scalable.
\end{lemma}
{\bf Proof: }
Assume that 
$\int_0^yk(x,y)\, dx=1$, $k(x,y)=k(y-x,y)$, and 
$ k(x,y) = k(x/y, 1)/y$. Then, 
$$
\int_0^y x^ik(x,y)\, dx = \int_0^y x^ik(x/y,1)/y\, dx = y^i\int_0^1\xi^i k(\xi,1)\, d\xi.$$
We already know that $\alpha_0=2$ and $\alpha_1=1$ 
(equation~(\ref{k2integral}) resp.\ lemma~\ref{k2m1}). 
The monotonicity of $\alpha_i$ 
and the fact that the sequence tends to zero follows from
$\alpha_i = \int_0^1\xi^i k(\xi,1)\, d\xi$. Moreover, 
$$ \infty > \int_0^1\ln(1/(1-x))\,|k(x,1)|\, dx
= \int_0^1\sum_{i=1}^\infty \frac {x^i}{i} k(x,1)\, dx
= \sum_{i=1}^\infty\alpha_i/i.$$
\qed\par\medskip

In difference to non-scalable kernels,
scalable kernels distribute plasmids in a similar way  
to the daughter cells, independently on the amount of plasmids the mother cell contains. 
In the case of low copy plasmids, there are plasmid-distribution systems ensuring that all 
daughter cells receive at least one plasmid. If we have very few plasmids, the distribution law for plasmids may explicitly depend on the number of available 
plasmids in a non-scalable way. However, if there are more than only few plasmids present,  
we expect a scalable law to appear.  In particular,
as we consider the continuum limit for high copy plasmids, the assumption of scalable kernels seems to be reasonable.\\
We again indicate that the kernel $k(x,y)$ always inherits the symmetry $k(y,x)=k(x-y,x)$ as we do not distinguish between the two daughters. Unequal segregation can be recognized in biological sensible cases by bimodal, equal segregation by unimodal shapes of the kernel. 
\par\medskip

\begin{example} $k(x,y) = 2\,\delta(x-y/2)$, with
$$\alpha_i = \int_0^1x^ik(x,1)\, dx = 2^{1-i}.$$
In this case, necessarily both daughter cells obtain the
same number of plasmids. Plasmid segregation is necessarily
symmetric.
\end{example}
\begin{example} $k(x,y) = (2/y)\,\chi_{[0,y]}(x)$, with  
$$\alpha_i = \int_0^1x^ik(x,1)\, dx = \frac 2{i+1}.$$
\end{example}
\begin{example} 
Now we give an example for an asymmetric plasmid distribution between the daughters. One 
daughter  receives more plasmids, 
to be precise, 
$(1+a)y/2$ plasmids. Then, the other daughter cell receives 
$(1-a)y/2$ plasmids. This setup is modeled by the kernel 
$k(x,y) = \delta(x-(1-a)y/2)+\delta(x-(1+a)y/2)$, with 
$0\leq a < 1$ and 
$$\alpha_i = \int_0^1x^ik(x,1)\, dx = 2^{-i}\left\{(1-a)^i+(1+a)^i\right\}.$$
Note that the kernel $k(x,y)$ still possesses
a symmetric shape, though the underlying plasmid distribution mechanism is unsymmetrical. It is clear that there is also a 
symmetric plasmid segregation mechanism that yields the very same kernel.
\end{example}
\par\medskip

From these examples we conclude that an unequal segregation mechanism is 
likely to increase the variance in the kernel. This
observation allows to reformulate our initial problem: The question is, if a kernel with a high variance leads to 
accumulation of plasmids in a subpopulation, or if the variance has
only a minor effect on the plasmid distribution.\par\medskip

\begin{remark} 
 Prescribe a sequence $(a_i)_{i\in\N_0}$. The task to find 
a measure $\nu(x)$ with support $[0,1]$  such that  
$a_i=\int_0^1x^i\,d\nu(x)$ 
is called Hausdorff moment problem.
Necessary and sufficient conditions are known 
that guarantee a solution for the problem. 
Moreover, if a solution exists, 
it is unique~\cite{talenti1987}. However, 
the moment problem is ill posed, such
that naive numerical algorithms to reconstruct the measure from the moments are 
bound to fail. For us it is 
sufficient to note that the 
moments $\alpha_i$ defined above provide the complete 
information about a given, 
scalable kernel $k(x,y)$; in principle, it is possible to reconstruct $k(x,y)$ 
from the sequence $\alpha_i$.
\end{remark}

The next theorem is the central step to reformulate 
the dynamics in terms of the moments $M_i(t)=\int_0^\infty z^i u(z,t)\, dz$.  
In a similar spirit, Wake et al.~\cite{wake2014} consider a 
double Dirichlet series to investigate the fragmentation equation. 
The moment method is widely used in population genetics 
(see any text book about population genetics, e.g.\ Tavar\'e~\cite{tavare:popGen} or Durett~\cite{durett:popGen}), 
and we will find that it also yields useful results for the problem
addressed in the present paper.

\begin{theorem} \label{momentTheo} Let 
$z^iu(z,t)\in L^1$, $z^{i-1}b(z)u(z,t)\in L^1$, and 
 $M_i(t)=\int_0^\infty z^i u(z,t)\, dz$. Assume furthermore that the kernel $k(x,y)$ is scalable with moments $\alpha_i$. Then, 
\begin{eqnarray*}
M_0'(t) & = & 
(\beta-\mu)M_0(t) \\
M_1'(t) & = &   \int_0^\infty b(z)u(z,t)\, dz 
-\mu\,M_1(t) \\
M_i'(t) & = &  i\, \int_0^\infty z^{i-1}b(z)u(z,t)\, dz 
-(\beta\,(1-\alpha_i) + \mu)M_i(t) \quad\mbox{ for } i> 1.
\end{eqnarray*}
\end{theorem}
{\bf Proof:}
Multiplying equ.~(\ref{mod0X3}) by $z^i$ and integrating over $z$ yields
\begin{eqnarray*}
&&\frac d{dt} M_i(t)  =  \int_0^\infty z^i u_t(z,t)\, dz \\
&=& 
-\int_0^\infty z^i \partial_z(b(z) u(z,t)) dz -(\beta+\mu) \int_0^\infty z^i\,u(z,t)\,dz\\
&&\qquad\qquad\qquad+ \beta\, \int_0^\infty z^i\int_{z}^\infty k(z,z') u(z',t)\, dz'\,dz\\
&=& -\int_0^\infty z^i(b(z)u(z,t))_z\, dz 
-(\beta+\mu)M_i(t) + \beta\int_0^\infty \int_0^{z'} z^ik(z,z')\, dz\, u(z',t)\,dz' \\
&=& -\int_0^\infty z^i(b(z)u(z,t))_z\, dz 
-(\beta+\mu)M_i(t) + \beta\int_0^\infty  (z')^i\alpha_i\, u(z',t)\,dz' \\
&=& - z^ib(z)u(z,t)\,|_{z=0}^\infty  + i\, \int_0^\infty z^{i-1}b(z)u(z,t)\, dz
-(\beta (1-\alpha_i)+\mu)M_i(t). 
\end{eqnarray*}
The result follows with $b(0)u(0,t)=0$, $z^iu(z,t)\in L^1$, $\alpha_0=2$, and $\alpha_1=1$.
\qed\par\medskip
Note that the model (\ref{mod0X3}) preserves positivity, and hence all moments
are non-negative if we start with a non-negative initial condition $u(z,0)$.\\
The equation for $M_0(t)$ -- the total 
bacterial population size -- decouples 
from all higher moments. Basically, this 
finding is equivalent with  proposition~\ref{popDyn0}. We state this 
result again, this time in terms of $M_0(t)$.
\begin{proposition}\label{popDyn1}
$M_0(t) = e^{(\beta-\mu)\, t} M_0(0).$
\end{proposition}
We use theorem~\ref{momentTheo} as the starting point to 
 investigate the consequence of certain plasmid reproduction characteristics 
 by different choices of $b(z)$
for the plasmid dynamics. We discuss four cases:\par\medskip
\begin{tabular}{lll}
(a)& cell controlled mode& $b(z)=b_0$\\
(b)& plasmid controlled mode& $b(z)=b_0\,z$\\
(c)& cell controlled mode with carrying capacity& $b(z)=b_0\,(1-z/z_0)$.\\
(d)& plasmid controlled mode with carrying capacity   &$b(z)=b_0\,z(1-z/z_0)$\\
&(logistic reproduction)&
\end{tabular}

\subsection{Plasmid- and cell controlled mode 
            without carrying capacity}\label{simpleCase}
The cases investigated here give some general ideas about the long term behavior of 
the plasmid distribution; in particular, we develop 
ideas under which conditions the plasmids are lost
by the population. We will use these ideas when we analyze 
logistic and cell controlled 
reproduction with carrying capacity in the next section.
\par\medskip

\subsubsection{Cell controlled mode}
Let us assume that each cell produces plasmids at a constant rate, $b(z)= b_0$. We find  
$$ M_1' =  b_0 M_0 - \mu M_1.$$
Recall that $M_1(t)$ does not denote the number of plasmids per cell,
but the amount of plasmids in the total population. 
Asymptotically, we have $M_1(t)\sim e^{(\beta-\mu) t}$. 
The number of cells and the number of plasmids eventually grow with the same exponent. As
$$\left(\frac{M_1(t)}{M_0(t)}\right)' = b_0 - \beta \left(\frac{M_1(t)}{M_0(t)}\right)$$
the number of plasmids per cell tends to $b_0/\beta$. 
It does not grow unlimited as cell division is 
in the present case effective enough to control the number of plasmids. 
This is a potential difference to the next case. 
Note that existence of an exponential solution 
$u(z,t) = e^{(\beta-\mu)t}U(z)$ 
is well known for this case~\cite{perthame:book}.
\par\medskip
\noindent

\subsubsection{Plasmid controlled mode}
Every single plasmid replicates at rate $b_0$, such that $b(z)=b_0 z$; even 
if a cell already contains many plasmids, the reproduction rate 
of a single plasmid is not decreased. This is a linear model in all aspects (cell replication 
as well as plasmid replication),
$$M_1' = b_0 M_1-\mu M_1.$$
The dynamics of $M_0$ and $M_1$ decouple. 
We have  $M_1(t) = M_1(0) e^{(b_0-\mu) t}$ and the average number of plasmids per cell
is given by
$$ M_1(t)/M_0(t) = M_1(0)/M_0(0)\,\, e^{(b_0-\beta) t}.$$

\begin{corollary} Let $b(z) = b_0\, z$. 
Then
$ M_1(t)/M_0(t) \rightarrow 0$ if $b_0<\beta$, 
and $M_1(t)/M_0(t)~\rightarrow~\infty$ if 
$b_0>\beta$. 
\end{corollary}
This is a typical dichotomy we often find in epidemic models. 
We clearly see the consequence of the race between plasmid 
and bacterial reproduction visualized in Figure~\ref{model}. The 
plasmids reproduce at rate $b_0$, and 
their number increase exponentially fast. 
Cell divisions distribute the plasmids
to several cells, and decrease the number of plasmids per cell. 
In a thought experiment, we start with one single cell containing one single plasmid and
disregard cell death and stochastic effects. Since the 
plasmid reproduction rate is constant per plasmid (and does not 
depend on the number of plasmids in a cell), after time $t$ 
the number of all plasmids that are descendants of this primary 
plasmid (in all cell) is given by $\exp(b_0 \,t)$. The number of cells that are descendants of this first cell at
time $t$ reads $\exp(\beta \,t)$. Hence, the average number 
of plasmids per cell in this sub-population is $\exp((b_0-\beta) t)$.
As cell death affects bacteria and plasmids in the same way, 
it cancels out. The faster reproduction rate wins the race. 
There is no mechanism to stop the number  of plasmids per cell 
to go to infinity if $b_0>\beta$. This will be different in the next
section, where we incorporate a carrying capacity for plasmids in a cell. Note that the non-existence of an exponentially growing 
solution with a stable shape for the present case 
is already mentioned in~\cite{doumic2010}.
\par\medskip

\subsection{Plasmid- and cell controlled mode with carrying capacity}
\subsubsection{ Logistic production of plasmids}
We proceed to a more realistic scenario: 
we assume logistic growth for the plasmids, 
$$b(z) = b_0\, z(1-z/z_0).$$
For the present case, the existence of an asymptotic plasmid distribution
$U(z)$ seems not to be established by
now for $\beta$ constant. There are results for the case that $b(z)$ is logistic with $k\in L^{\infty}(\mathbb{R}^2_+)$ and $\beta(0)=0$~\cite{doumic2007}, but the case that $k$ is not bounded and $\beta$ is a positive constant seems not to be considered.
\par\medskip

\begin{figure}[htb]
\begin{center}
\includegraphics[width=6cm]{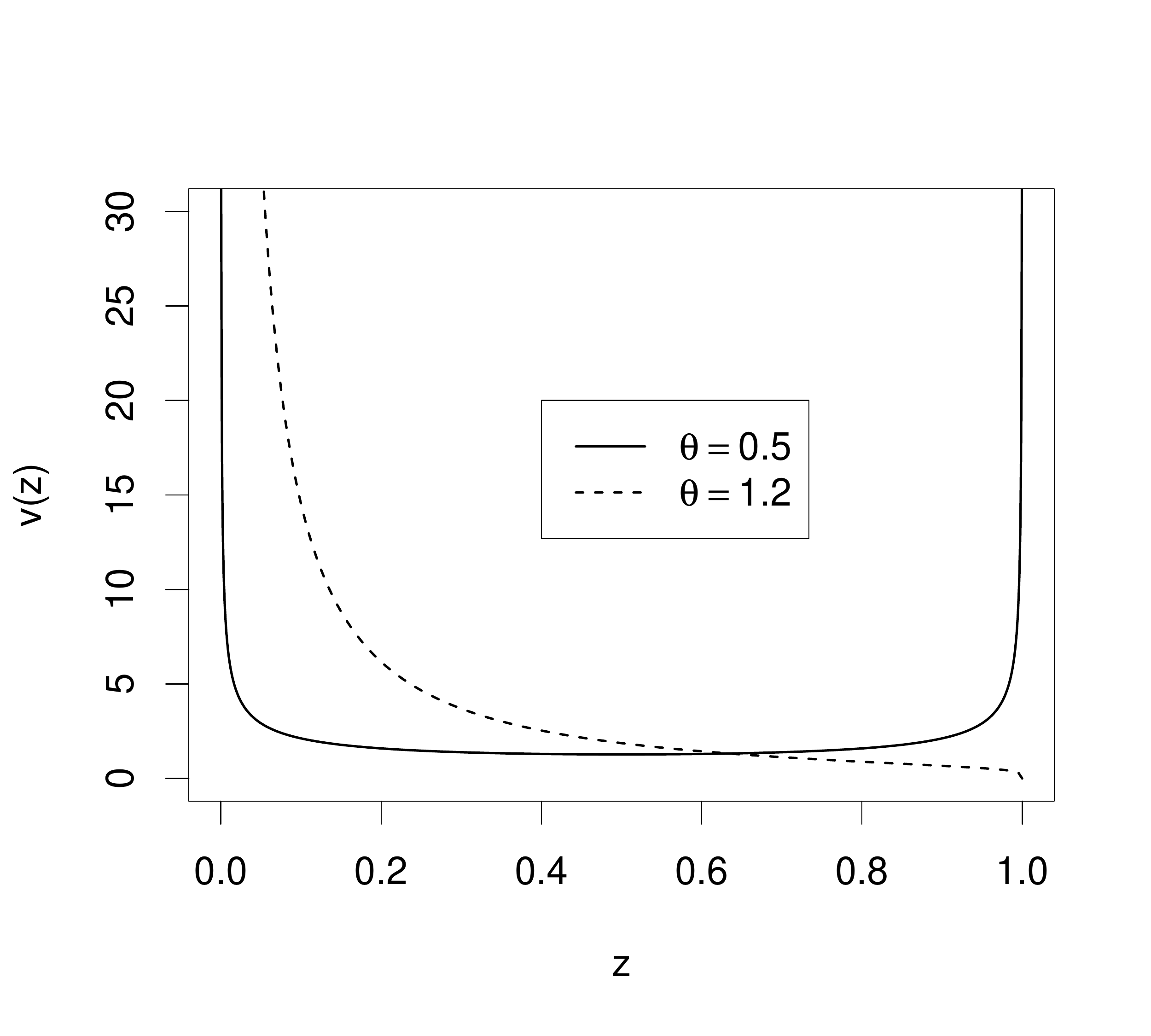}
\end{center}
\caption{Logistic plasmid reproduction and uniform distribution for $\theta=2\beta/b_0=0.5$ (solid line) and $\theta=1.2$ (dashed line). The functions are normalized such that their first moment agree.}\label{logGrowPic}
\end{figure}

Before we start with general considerations, 
it is instructive to investigate a situation where it is possible to 
explicitly compute $U(z)$:
If we 
use $k(x,y)=1/y$ as kernel, 
the function $U(z)$ satisfies the integro-differential equation
\begin{eqnarray*}
(b_0 z (1-z/z_0))U(z))_{z} + 2\beta U(z)
-2\beta\int_z^{z_0}k(z,z')U(z')\, dz' &=&0\\
\Rightarrow \quad
(z (1-z/z_0))U(z))_{zz} + \theta U_z(z)
+\theta U(z)/z &=&0
\end{eqnarray*}
with $\theta=2\beta/b_0$. We take $z_0=1$. There is an explicit solution
of this linear ordinary differential equation, given by
$U(z) 
 =  A\, z^{-1} + B\, z^{-\theta}\,(1-z)^{\theta-1}.
$ 
The first part of the fundamental solution does not satisfy the integro-differential equation, the second part well. We have 
$$U(z) = z^{-\theta}\,(1-z)^{\theta-1}.$$ 
For $\theta\geq 1$, the integrability of this function at $z=0$ 
is not given, such that 
the zero'th
moment only exists for $\theta<1$. It is interesting
to note that we always have a pole at $z=0$.
This is a fundamental difference to the cell controlled plasmid production modes. 
For $\theta<1$, we find a bimodal distribution -- cells tend to have either very few or many plasmids (see Figure~\ref{logGrowPic}). 
We may interpret this finding again
in terms of the race between cell- and plasmid reproduction: for 
$z$ small, the reproduction of plasmids is outraced by the reproduction of cells. Therefore, cells with only few plasmids are washed towards $z=0$. If $z\approx 1$, the situation reverse: the plasmids 
reproduce much faster than the cells divide, and hence there the cells
are driven towards large plasmid numbers. This mechanism could be used
by a population to create subpopulations with 
few resp.\ many plasmids, fulfilling different tasks. 
Such division of labor is well known to be optimal in 
the context of switching environments (bet hedging, 
see e.g.~\cite{acar,mueller:bedHedge}). 
Depending on the properties the plasmid codes, this division of 
labor could be of advantage for the complete population.\\ 
As our interest 
is primarily the question if plasmids accumulate within cells, 
we investigate if the behavior of our example at $z=z_0$
is special or generic.
\par\medskip

With the present choice of $b(x)$, we obtain 
a hierarchical system for the moments
\begin{eqnarray}
M_0'(t) & = & (\beta-\mu)\, M_0(t)\label{momentSytsLogist1}\\
M_1'(t) & = & (b_0-\mu) M_1(t) - b_0/z_0 M_2(t)\\
M_i'(t) & = & (i\,b_0-(1-\alpha_i)\beta-\mu) M_i(t) - i\,b_0/z_0 M_{i+1}(t) \quad \mbox{ for } i>1.\label{momentSytsLogist3}
\end{eqnarray}
Note that $\alpha_1=1$, s.t.\ the second equation is a special case of the third equation. 
We expect for very few plasmids that $M_i$ is dominating $M_{i+1}$ 
(take, for example,  $u(z,0)\approx \delta_{\varepsilon}(z)$, then $M_i(0)\approx \varepsilon^i$). 
Therefore, we expect in particular $M_1$ to grow with exponent $b_0-\mu$. 
This growth is eventual stopped if the plasmids spread and $M_2$ starts to grow.
The structure we find here reminds of the linerization 
at the uninfected solution for a model for infectious diseases.
\par
Asymptotically, it is likely that either  the average
number of plasmids per cell $M_1/M_0$ tends to zero (if $b_0<\beta$),
or becomes constant (if $b_0>\beta$). The next proposition supports the first idea.           
\par\medskip

\begin{theorem} {\bf [Threshold Theorem]} If $b_0<\beta$ then $M_1(t)/M_0(t)\rightarrow 0$ for $t\rightarrow\infty$.
\end{theorem}
{\bf Proof:} Since $M_1'\leq~(b_0-\mu)~M_1$, we have the upper bound $~M_1(t)\leq~M_1(0)~e^{(b_0-\mu)t}$. 
And as $M_0(t)~M_0(0)~e^{(\beta-\mu)t}$ the result follows.\qed
\par\medskip

Now we turn to the case $b_0>\beta$: 
If we inspect the equation for $M_i'(t)$, the constant in front of $M_i$ 
is $i\,b_0-(1-\alpha_i)\beta-\mu$. 
Since $\alpha_i\rightarrow 0$ for $i\rightarrow \infty$, 
this term tends to infinity. One could think that the higher moments grow exponentially at a rate constant that is arbitrarily large.
 The following proposition shows that this is not the case. 

\begin{proposition} \label{momentGrowthPos}Let $u(z,0)$ satisfy the 
conditions of lemma~\ref{finiteSupport}.
Let furthermore $\beta<b_0$, $\xi_1=1$, and 
$$ \xi_{i} = 
   \xi_{i-1} \,\,\frac{i-1}{z_0\,(i-1-(1-\alpha_{i})\,\beta/b_0)}.$$
Then, $\xi_i>0$, and 
   $$ \frac d {dt} \left(\sum_{i=1}^\infty \xi_i M_i\right) = (b_0-\mu) \,  \left(\sum_{i=1}^\infty \xi_i M_i\right).$$
   Moreover, 
   $$M_i(t)\leq z_0^i M_0(0)\, e^{(\beta-\mu)\, t}.$$
\end{proposition}
{\bf Proof: } Since the conditions of Lemma~\ref{finiteSupport} are given, we know 
that for any time $t$ there is $\varepsilon(t)>0$ such that 
$\supp(u_0)\subset[0,z_0-\varepsilon(t)]$. 
Hence, 
$$M_i=\int_0^\infty z^iu(z,t)\, dz = 
\int_0^{z_0-\varepsilon(t)} z^iu(z,t)\, dz  \leq (z_0-\varepsilon(t))^i\, M_0(t)$$
which implies the upper bound for $M_i(t)$. Furthermore, 
the sequences we discuss in 
this proof do all converge uniformly. In particular,
we are allowed to exchange the infinite sum 
and the derivative.

\begin{eqnarray*}
\frac d {dt}\sum_{i=1}^\infty \xi_i\, M_i
&=& b_0 \sum_{i=1}^\infty \xi_i\, \big[(i\,-(1-\alpha_i)\beta/b_0-\mu/b_0) M_i - i/z_0 M_{i+1}\big]\\
&=& b_0\, (1-(1-\alpha_1)\beta/b_0) M_1 +  b_0 \sum_{i=2}^\infty \xi_i\, (i-(1-\alpha_i)\beta/b_0) M_i\\
&&\qquad -  \mu \sum_{i=1}^\infty \xi_i M_{i}-  b_0 \sum_{i=1}^\infty \xi_i(i/z_0) M_{i+1} 
\end{eqnarray*}
At this point,  we use $\alpha_1=1$ and proceed
\begin{eqnarray*}
&& b_0\, M_1 +  b_0 \sum_{i=2}^\infty \xi_i\, (i-(1-\alpha_i)\beta/b_0) M_i -  b_0 \sum_{i=2}^\infty \xi_{i-1}((i-1)/z_0) M_{i} -  \mu \sum_{i=1}^\infty \xi_i M_{i}\\
&=& b_0\, M_1 +  b_0 \sum_{i=2}^\infty \xi_i\, (i-(1-\alpha_i)\beta/b_0) M_i -  b_0 \sum_{i=2}^\infty \frac{(i-1)\,z_0\,(i-1-(1-\alpha_{i})\,\beta/b_0)    }{(i-1)\,z_0} \xi_iM_{i}\\
&&\qquad  -  \mu \sum_{i=1}^\infty \xi_i M_{i}\\
&=& b_0 M_1 +  b_0 \sum_{i=2}^\infty \xi_i\,\left[(i-(1-\alpha_i)\beta/b_0)-(i-1-(1-\alpha_{i})\,\beta/b_0)\right]M_i -  \mu \sum_{i=1}^\infty \xi_i M_{i}\\
&=& (b_0-\mu) \,  \left(\sum_{i=1}^\infty \xi_i M_i\right)
\end{eqnarray*}
\qed\par\medskip

Before we reformulate this result in terms of $u(z,t)$, we state a 
simple result about the asymptotic behavior of a sequence constructed 
in a similar way as $\xi_i$.
\begin{lemma} \label{folgenAsmpytLemma} Let $a>0$, $b_i\geq 0$ with $\sum_{i=1}^\infty b_i/i<\infty$, and $1-a/i+b_i/i>0$ for  all $i\in\N$, 
$$y_1=1,\quad y_{i+1}=\frac{y_i}{1-a/i+b_i/i}.$$ 
Then, there are 
$c_1, c_2\in \R$, $c_1,c_2>0$ such that $c_1 i^a \leq y_i\leq c_2 i^a$.
\end{lemma}
{\bf Proof: } 
Let $z_i=i^{-a} y_i$, then 
$z_{i+1}/{z_i} 
= (1+1/i)^{-a}\,\,(1-a/i+b_i/i)^{-1}$
and 
$$ -\ln(z_i) 
= \sum_{\ell=1}^{i-1} \bigg[
a\,\ln(1+1/\ell) + \ln(1-a/\ell+b_\ell/\ell)\bigg]
.$$
We show that there is a uniform upper and lower bound for the sum at the r.h.s.\\
Lower bound: Since $a\,\ln(1+1/\ell) + \ln(1-a/\ell+b_\ell/\ell)\geq 
a\,\ln(1+1/\ell) + \ln(1-a/\ell)$ for $\ell$ sufficiently large to ensure that $1>a/\ell$, we define $g(a,x) = a\ln(1+x)+\ln(1-ax)$ and study $\sum_{\ell=\ell_0}^{i-1} g(a,1/\ell)$. 
Choose $\ell_0>10+1/a$. As $g(a,0)=\partial_x g(a,0)=0$, we find $c>0$ such that $|g(a,x)|\leq c x^2$ for $0\leq x<1/\ell_0$. Then, for $i>\ell_0$, 
$$\sum_{\ell=1}^{i-1} g(a,1/\ell) = \sum_{\ell=1}^{\ell_0} g(a,1/\ell)
+ \sum_{\ell=\ell_0+1}^{i-1} g(a,1/\ell) \geq  \sum_{\ell=1}^{\ell_0} g(a,1/\ell)- c  \sum_{\ell=\ell_0+1}^{\infty}\ell^{-2}$$
and $\sum_{\ell=1}^{i-1} \bigg[
a\,\ln(1+1/\ell) + \ln(1-a/\ell+b_\ell/\ell)\bigg]$ is uniformly bounded from below.\\
Upper bound: We have for $i>\ell_0$ 
\begin{eqnarray*}
 \sum_{\ell=\ell_0}^{i-1} \bigg[
a\,\ln(1+1/\ell) + \ln(1-a/\ell+b_\ell/\ell)\bigg]
\leq \sum_{\ell=\ell_0}^\infty g(a,1/\ell) 
+ \sum_{\ell=\ell_0}^\infty  \ln\left(1 + \frac{b_\ell/\ell}{1-a/\ell}\right).\end{eqnarray*}
The first sum on the r.h.s. is finite with the same argument we used 
above. Since for $\ell$ large, 
$\ln\left(1 + \frac{b_\ell/\ell}{1-a/\ell}\right)\leq c\, b_\ell/\ell$ 
for some $c>1$,  
the condition $\sum_{i=1}^\infty b_i/i<\infty$ implies that also the second sum is bounded.
\par\qed\par\medskip

\begin{proposition} \label{corGrwoth}Let $\varphi(z) = \sum_{i=1}^\infty \xi_i z^i$. 
 If $b_0>\beta$, the convergence radius of this power series
is $z_0$. Moreover, $\varphi(z)\rightarrow\infty$ for $z\rightarrow z_0-$. 
Let $\supp(u(z,0)) \subset[0,z_0-\varepsilon]$, $\varepsilon>0$. Then we find
$$ \frac d {dt}
\int_0^{z_0} \varphi(z) u(z,t)\, dz
=  (b_0-\mu)\,\int_0^{z_0} \varphi(z) u(z,t)\, dz.$$
\end{proposition}
{\bf Proof: }
Since $\alpha_1=1$, and the sequence $\alpha_i$  
is strictly monotonously decreasing, 
and therefore $\xi_i>0$ 
 in case of $\beta\leq b_0$. As 
 $$ \xi_{i} = 
    \,\,\frac{\xi_{i-1}\,(i-1)}{z_0\,(i-1-(1-\alpha_{i})\,\beta/b_0)}
= 
    \,\,\frac{\xi_{i-1}}{z_0\,(1- (\beta/b_0)\,/(i-1) + (\beta/b_0)\,\alpha_{i}/(i-1))}   
   $$
 the asymptotical behavior of 
 $\xi_i$ can be determined via Lemma~\ref{folgenAsmpytLemma} 
 (note that our definition of scalable kernels implies that 
 $\sum\alpha_i/i<\infty$); we find $c_1,c_2>0$ such that
$$ c_1  \,\, z_0^{-i} i^{\beta/b_0} > \xi_i > c_2  \,\, z_0^{-i} i^{\beta/b_0}.$$
Hence $\varphi(z)$ is analytic in the complex circle 
$\{|z|<z_0\}$, and $\varphi(z)\rightarrow\infty$ for $z\rightarrow z_0-$. As (again, lemma~\ref{finiteSupport}
guarantees the proper convergence)
$$ \sum_{i=1}^\infty \xi_i M_i(t) = \int_0^{z_0} \left(\sum_{i=1}^\infty \xi_i z^i\right) u(z,t)\, dz
= \int_0^{z_0} \varphi(z) u(z,t)\, dz
$$
we find with the help of proposition~\ref{momentGrowthPos}  the desired ordinary differential equation.\qed\par\medskip

This proposition gives a first hint about the behaviour of $U(z)$ for $z\rightarrow z_0-$: 
The weighted sum of the moments tends exponentially 
fast to infinity with exponent $b_0-\mu$.  
We have the bound $M_i\leq z_0^iM_0(0)e^{(\beta-\mu) t}$ for each moment and in case $\beta<b_0$. Hence,  
we expect that $u(z,t)$ tends to a solution 
$U(z)e^{(\beta-\mu) t}$ 
where $\int_0^\infty U(z)\varphi(z)\, dz = \infty$. 
This proposition indicates that, even if $U(z)$ 
tends to zero for $z\rightarrow z_0-$, this function 
must not decline too fast. 
We utilize the system of ordinary differential equations  
(\ref{momentSytsLogist1})-(\ref{momentSytsLogist3})  to obtain an idea how $U(z)$ may look like.

\begin{proposition}\label{xxxProp}
Assume that $u(z,t)=e^{(\beta-\mu) t} U(z)$ with $ z^{\ell_0}U(z)\in L^1(0,z_0)$ 
for some $\ell_0\in\N_0$. Let $P_i=\int_0^\infty z^i U(z)\, dz$  for $i\geq\ell_0$. Then
$P_i=P_{\ell_0}\,\eta_i$ for $i>\ell_0$ with 
$$\eta_i = z_0^{i-\ell_0}\prod_{j=\ell_0}^{i-1}\left(1-\frac{(2-\alpha_j)\beta}{j\,b_0}\right). $$
\end{proposition}
{\bf Proof: }
Note that $z^i U(z)\in L^1(0,z_0)$ if $z^{\ell_0}U(z)\in L^1(0,z_0)$ and $i\geq \ell_0$. 
If $u(z,t)=e^{(\beta-\mu) t} U(z)$ is true, then
$M_i(t)=e^{(\beta-\mu)t} P_i$, and due to equ.~(\ref{momentSytsLogist3})
$$(\beta-\mu) P_i =  (i\,b-(1-\alpha_i)\beta-\mu) P_i - i\,b_0/z_0 P_{i+1}.$$
Hence,
$$P_{i+1} = z_0\,\left(1-\frac{(2-\alpha_i)\beta}{i\,b_0}\right)\, P_i.$$
\qed\par\medskip

\begin{proposition}
Let $P_i$ defined as in proposition~\ref{xxxProp}. 
There are constants $c_1,c_2>0$ such that 
$$ c_1 \, \,i^{-2\beta/b_0}\,\, \leq P_i/z_0^{i-1} \leq c_2\,\,i^{-2\beta/b_0}.$$
\end{proposition}
{\bf Proof:} The result is a consequence of Lemma~\ref{folgenAsmpytLemma}, as 
$k(x,y)$ is assumed to be a scalable kernel, and  
$$\left( \prod_{j=\ell_0}^{i-1}\left(1-\frac{(2-\alpha_j)\beta}{j\,b_0}\right)\right)^{-1} 
= \prod_{j=\ell_0}^{i-1}\left(
\frac{1}
     {1-(2\beta/b_0)/j+(\alpha_j\beta/b_0)/j}\right).
$$
\qed\par\medskip

The asymptotics of the moments give some hint about the shape of $U(z)$ 
for $z$ close to $z_0$, as $(z/z_0)^i$ tends point wise to zero for $z<z_0$ and $i\rightarrow \infty$. 
It is instructive to compare with a function $V(z)$ with support in $[0,1]$
(take $z_0=1$) and moments $i^{-\theta}$. 

\begin{remark}
The function 
$$ V(z) = \ln(1/z)^{\theta-1}/\Gamma(\theta),\qquad z\in(0,1), \quad \theta>0$$ 
has moments $\int_0^1 z^i V(z)\, dz = (i+1)^{-\theta}$ for $i\in\N_0$ 
(see e.g.~\cite{bdur} or \cite[p.~550, 4.272, 6.]{GR}). Then, $\lim_{z\rightarrow 0} V(z)=0$
for $\theta\in(0,1)$, $V(1)=1$ if $\theta=1$, and $V(z)\rightarrow\infty$ for $z\rightarrow 1-$ in case of $\theta>1$.
\par\medskip
Heuristically,  we identify $\theta= 2\beta/b_0$ and expect a similar behavior
for $U(z)$ at the right hand side of $[0,z_0]$ like $V(z)$ at $z=1$. In particular  we expect 
that the asymptotics of $U(z)$ for $z\rightarrow z_0$ dramatically changes 
at $b_0= 2\beta$. We may even expect that the asymptotics of $U(z)$ and $V(z)$ are similar. 
The following proposition supports this idea; 
we have seen this effect before in the introductory, explicit example 
at the beginning of this section.
\end{remark}

\begin{theorem}\label{asymptTheo}
Assume that there is $\delta>0$ 
such that $U|_{[z_0-\delta,z_0)}$ is continuous and monotonic. 
Then, $U(z)\rightarrow 0$ for $z\rightarrow z_0$ if $b_0\in(\beta,2\beta)$, and $U(z)\rightarrow\infty$ for $z\rightarrow z_0$ if $b_0>2\beta$.
\end{theorem}
{\bf Proof: } Without restriction we take $z_0=1$. Assume that $b_0 \in (\beta,2\beta)$, but $U(z)\not\rightarrow 0$ 
for $z\rightarrow 1$. As $U(z)$ is monotonic in $[1-\delta, 1)$, we have $U(z)>c>0$ 
within this interval, and 
$$ P_i \geq \int_{1-\delta}^1 x^i c \, dx = \frac{c}{1+i}\left(1-(1-\delta)^{i+1})\right)
\quad\Rightarrow\quad
\liminf_{i\rightarrow\infty}\,(i\, P_i)\geq c>0.
$$
As $P_i={\cal O}\left(i^{-2\beta/b_0}\right)$, we know that 
(note $2\beta/b_0>1$ if $b_0 \in (\beta,2\beta)$)
 $$\lim_{i\rightarrow\infty}i\,P_i
= \lim_{i\rightarrow\infty} \left(P_i i^{2\beta/b_0}\right)  i^{-2\beta/b_0+1} = 0.$$
 We obtain a contradiction, and hence
$U(z)\rightarrow 0$ for $z\rightarrow 1$.\\
For the case $b_0>2\beta$, we use a similar argument. If $U(z)\leq c$ in $[1-\delta,1]$, 
then
$$ P_i \leq \int_0^{1-\delta} x^iU(x)\, dx + c\, \int_{1-\delta}^1 x^i\, dx
\leq (1-\delta)^i P_0 +c \frac {1-(1-\delta)^{i+1}}{1+i}$$
and hence
$\limsup_{i\rightarrow\infty} i\, P_i \leq c$. However, we have 
in the present case $1-2\beta/b_0>0$, and thus
$$ \limsup_{i\rightarrow\infty} i\, P_i = 
\limsup_{i\rightarrow\infty} \left(P_i i^{2\beta/b_0}\right)  i^{-2\beta/b_0+1} = \infty.$$
\qed\par\medskip
\noindent
We combine this and the threshold theorem in the next corollary.

\begin{corollary} 
{\bf Case 1:} $b_0<\beta$. Then, 
 $M_1(t)/M_0(t)\rightarrow 0$ for $t\rightarrow\infty$, 
 i.e., plasmids are lost.\\
 For case~2 and case~3 assume that there is $\delta>0$  such that
  here is a solution $u(z,t) = c\,e^{(\beta-\mu) t}U(z)$ and  $U(z)|_{[z_0-\delta,z_0)}$ 
is continuous and monotonic. \\
{\bf Case 2:} $\beta < b_0<2\beta$. Then, $U(z)\rightarrow 0$ for $z\rightarrow z_0-$, 
i.e.\ plasmids do not accumulate at $z_0$.\\
{\bf Case 3:} $2\beta<b_0$. Then, $U(z)\rightarrow \infty$ for $z\rightarrow z_0-$.
 That is, plasmids accumulate at $z_0$.
\end{corollary}

\begin{remark}
 Note that the characteristics of the 
transition kernel $k(x,y)$ do not influence 
at all the thresholds stated in the corollary.
However, also the segregation characteristics affects 
the shape of the equilibrium distribution. 
Assume that $P_i$ and $\tilde P_i$ are moments connected with kernels 
moments $\alpha_i$ and $\tilde\alpha_i$. Then,
$$P_i>\tilde P_i\quad\mbox{ if } \alpha_i\geq \tilde\alpha_i 
\mbox{ and } P_{\ell_0}=\tilde P_{\ell_0}.$$
Heuristically, 
a kernel that describes an unequal plasmid transition has larger moments $\alpha_i$, $i>2$, than
an symmetric kernel.  
E.g., for $k(x,y) = \delta(x-y/2)$ we have 
$$\alpha_i = 2^{1-i}$$
while for 
$k(x,y) = \delta(x-(1-a)y/2)+\delta(x-(1+a)y/2)$, with 
$0< a < 1$, we obtain
\begin{eqnarray*}
\tilde \alpha_i& =& \int_0^1x^ik(x,1)\, dx 
= 2^{-i}\left((1-a)^i+(1+a)^i\right)\\
&=& 2^{-i}\left(\sum_{n=0}^i{ i \choose n}(1+(-1)^n)a^n\right)
> 2^{1-i}
\end{eqnarray*}
for $i>1$. The moments $\tilde \alpha_i$, $i>2$, are strictly monotonously increasing in $a$, and therefore, also the 
moments $\tilde P_i$. This indicates that the distribution 
$U(z)$ moves its maximum to the right, towards an accumulation of plasmids in some cells. However, the
kernel is never able to change the asymptotics of $U(z)$ 
for $z\rightarrow z_0-$.
\end{remark}

\subsubsection{Cell controlled mode with carrying capacity} 
In this last case, we assume that every cell replicates plasmids at 
constant rate $\beta$, but 
plasmid load reduces the reproduction rate, such that 
$b(z)=b_0 (1-z/z_0)$. 
Concerning the existence of a stable, 
asymptotic shape for the plasmid
distribution, there is some indication in the book of Perthame~\cite{perthame:book}, who considers $b(x)$ with a bounded 
support. There it is assumed that $b(x)$ is bounded away from zero on its support. Using perturbation methods, 
e.g.\ developed 
in~\cite{doumic2010}, it is most likely possible to establish existence also for our choice of $b(x)$. However, as stated before, we investigate the shape of the asymptotic plasmid distribution, and
take the existence for granted. \par 

 For the moments, we find the equations
$$M_i' = b_0 M_{i-1}-(\mu+\beta(1-\alpha_i)+b_0/z0) M_i\quad
\mbox{ for } i\geq 1.
$$
Obviously, there is no way for the population
to get rid of plasmids. 
\begin{proposition}
We find that for $i$ fixed
$$ M_i(t)/e^{(\beta-\mu)\,t}\rightarrow P_i$$
where $P_i$ are constants that satisfy for $i_0>0$
$$ P_{i+1}\, z_0 = \frac{1}{1+z_0\frac{\beta(2-\alpha_i)}{i\, b_0}}\, P_i \qquad\mbox{ for } i > 1.
$$
 $\alpha_i$ denote the moments of 
a scalable kernel $k(x,y)$.
\end{proposition}
{\bf Proof: }
The asymptotic behavior of $M_i(t)$ follows from
$M_0\sim e^{(\beta-\mu)t}$ together with induction. 
Asymptotically, we have $M_i=e^{(\beta-\mu)\,t}P_i$. 
Plugging this formula into the ordinary differential equation for $M_i$, we obtain the recursion formula  for $P_i$
$$ (\beta-\mu) P_i = b_0 P_{i-1}-(\mu+\beta(1-\alpha_i)+b_0/z0) P_i,$$
which yields the representation of $P_i$. 
\qed\par\medskip

Note that we do not claim a uniform convergence 
of the moments, but only convergence for any moment with $i$ fixed. This difference may be of importance for the convergence of $u(z,t)$ for $t\rightarrow\infty$. 
An argument similar to Theorem~\ref{asymptTheo} 
yields some information about the asymptotic behavior of $U(z)$ defined by 
$u(z,t) \sim  e^{(\beta -\mu) t} U(z)$ for $z\rightarrow z_0$. 

\begin{proposition}
Assume that there is $\delta>0$ 
such that $U|_{[1-\delta,1)}$ is continuous and monotonic. 
Then, $U(z)\rightarrow 0$ for $z\rightarrow z_0$ if $b_0\in(\beta,2\beta\,z_0)$, and $U(z)\rightarrow\infty$ for $z\rightarrow z_0$ if $b_0>2\beta\,z_0$.
\end{proposition}
The behavior of $U(z)$ at $z=0$ is determined by the boundary condition
$b(0) u(0,t)=0$. We have $U(0)=0$. This is a central difference 
to the logistic case, where plasmid reproduction tends to zero for 
$z\rightarrow 0$. Therefore, in the logistic case, $U(z)$ is likely to blow up 
if $z$ tends to zero.

\begin{figure}[tbh]
\begin{center}
\includegraphics[width=\textwidth]{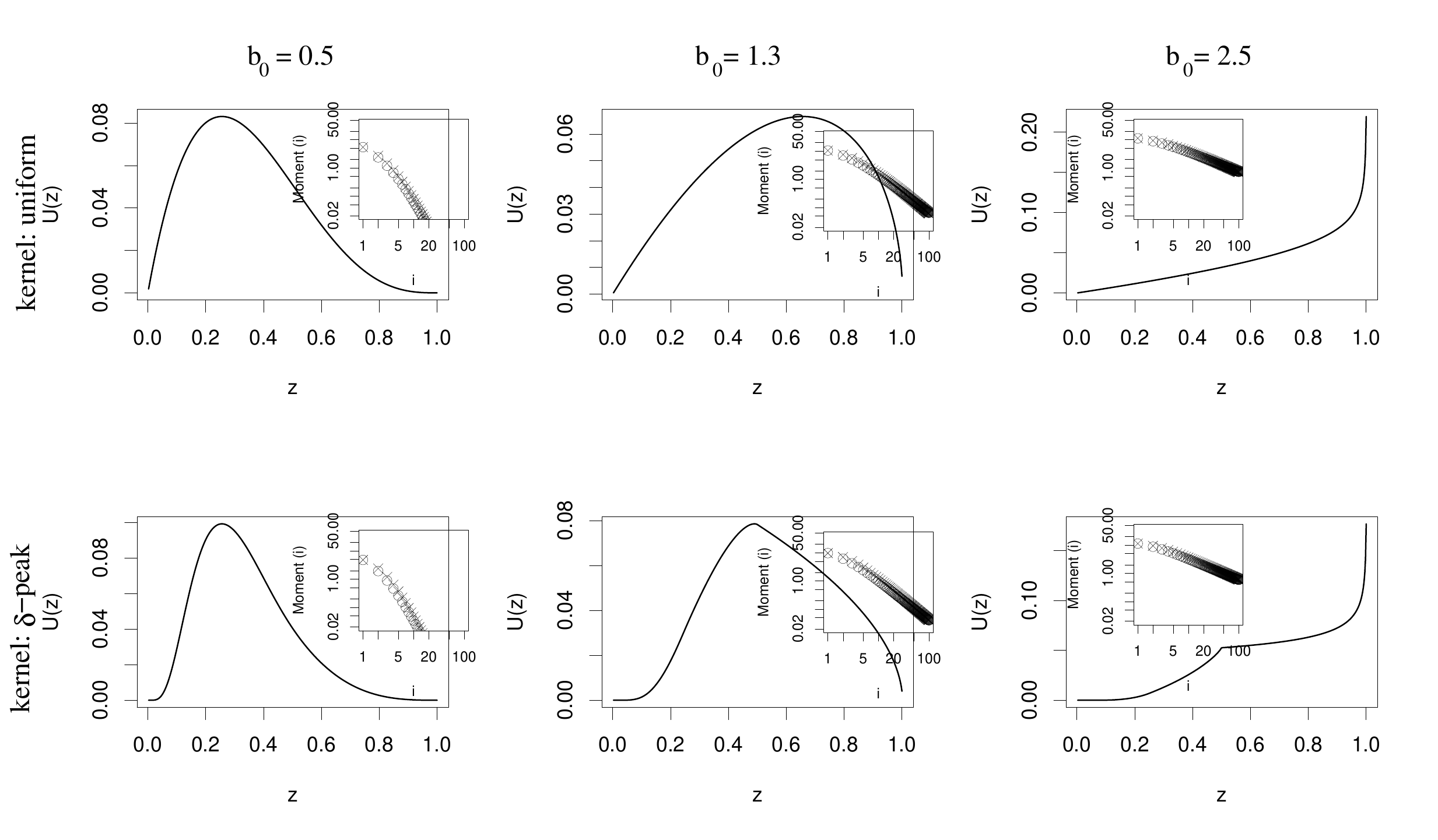}
\end{center}
\caption{Shape of equilibrium plasmid distribution 
for cell controlled plasmid production with
carrying capacity ($\beta=1.0$, $\mu=0$), uniform
transmission kernel (upper row) and the $\delta$-peak 
as transmission kernel (lower row); the three columns
correspond to the indicated value for $b_0$. The inlays show a comparison of the
theoretical moments (open bullets) and 
numerical moments (cross). 
}\label{symKenelPic}
\end{figure}

\section{Influence of model parameters on the  plasmid distribution}
\label{shapeSect}
We address here differences in the 
plasmid distribution caused by  different model parameters. 
We do know the moments of the stationary plasmid distribution, and hence, in theory it is possible to reconstruct this distribution.
However, as the 
Hausdorff moment problem is ill posed, 
we take another approach: We discretize 
the partial differential equation~(\ref{mod0X3}), 
i.e. return to equation~(\ref{discreteModelEq}).
This equation is  linear, i.e.,\ can be
written as $x'=A x$ where $A$ is a matrix and $x=(x_1,..,x_n)^T$ a vector. We then determine all
eigenvectors of $A$, and pick the appropriate positive
eigenvector. In case of cell controlled plasmid production, 
this eigenvector is unique due to the Perron-Frobenius theory; for the plasmid controlled case, the matrix $A$ is not irreducible, and two non-negative eigenvectors appear.
In order to check the numerical approach,
we compare the numerical moments of the resulting 
vector with the theoretical moments as computed above.
The result is displayed in Figure~\ref{symKenelPic}. 
We find, first of all, an excellent agreement between numerical and theoretical moments. 
We furthermore 
find hardly a difference between the moments for 
different kernels and the same dynamic parameters 
(distributions in the same column), but a distinct
difference between distributions for different kinetic parameters (distributions within one row). 
This weak
dependence on the specific plasmid transmission kernel is a consequence of the
ill posedness of the 
Hausdorff moment problem. However, 
the kernel ($\delta$-peak or uniform kernel) 
does have kind of second order an effect 
on the exact shape of the 
distribution, but the shape is by far more 
influenced by kinetic parameters 
(that is, by $b_0$ and $\beta$). 
\par\medskip

\begin{figure}[htb]
\begin{center}
\includegraphics[width=12cm]{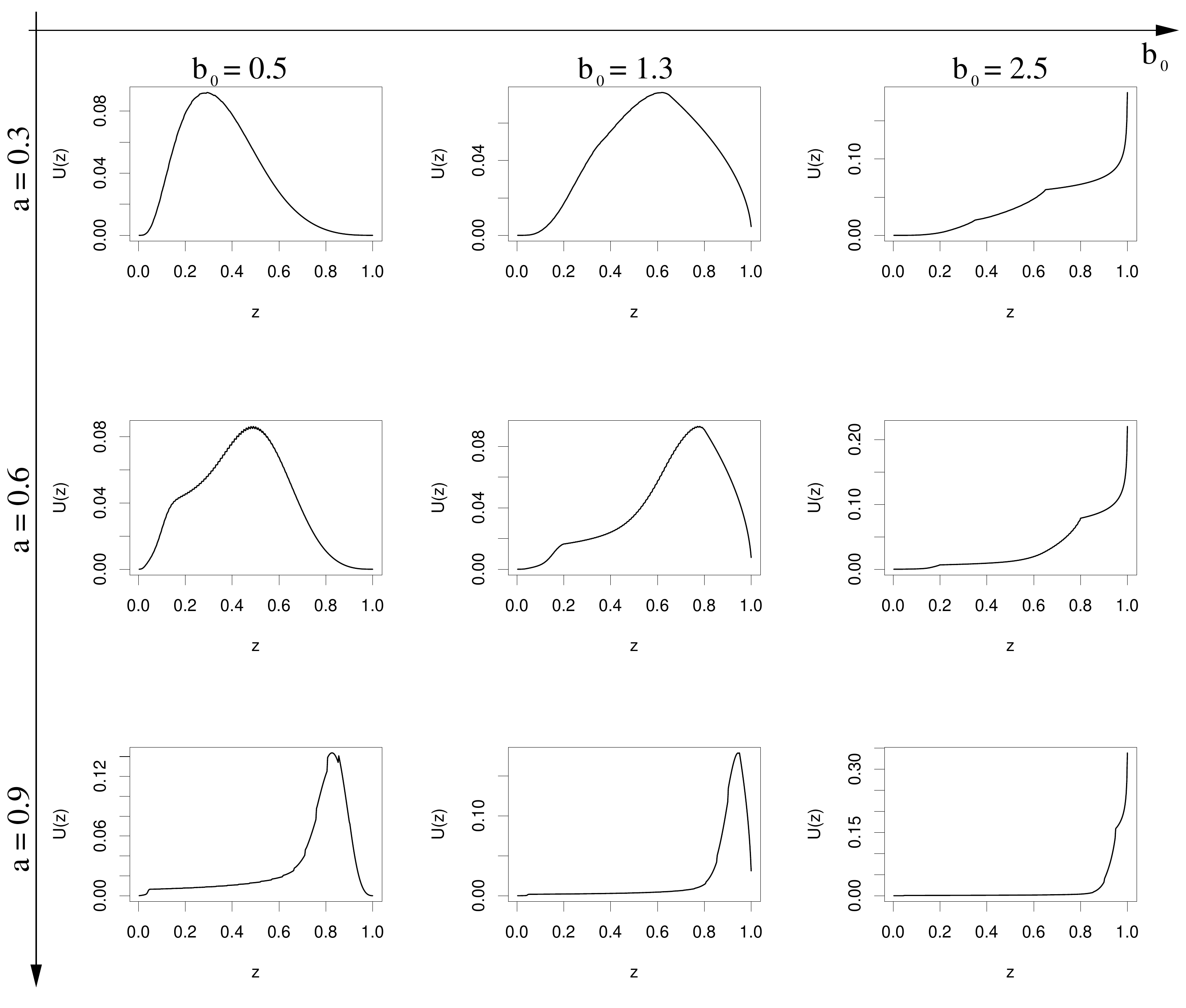}
\end{center}
\caption{Effect of unequal plasmid transmission 
(for cell controlled plasmid production with 
carrying capacity). Parameter $a\in (0,1)$ indicates the degree of unequality. $\beta=1$, $\mu=0$, $b_0$ as indicated.}\label{unequalPic}
\end{figure}

We visualize the effect of symmetric and 
non-symmetric
transmission characteristics between mother and daughter cells, 
or, equivalently, the effect of the 
variance of the fragmentation kernel. Therefore, we consider the kernel
$k(x,y) = \delta_{(1-a)y/2}(x) + \delta_{(1+a)y/2}(x)$, 
such that $a=0$ corresponds to a completely symmetric 
transmission, and $a=1$ a maximal non-symmetric transmission characteristics. 
We find in Figure~\ref{unequalPic} that non-symmetric 
plasmid transmission has its largest 
effect if the reproduction of plasmids $b_0$ is in 
the same range as the reproduction of cells. 
Most likely, this is the relevant case for many biological systems. In this range, a distinct 
non-symmetry is able to shift the peak of the distribution towards larger $z$-values, that is, 
cells tend to accumulate plasmids.

\section{Experimental findings}\label{experi}

\begin{figure}[htb]
\begin{center}
\includegraphics[width=12cm]{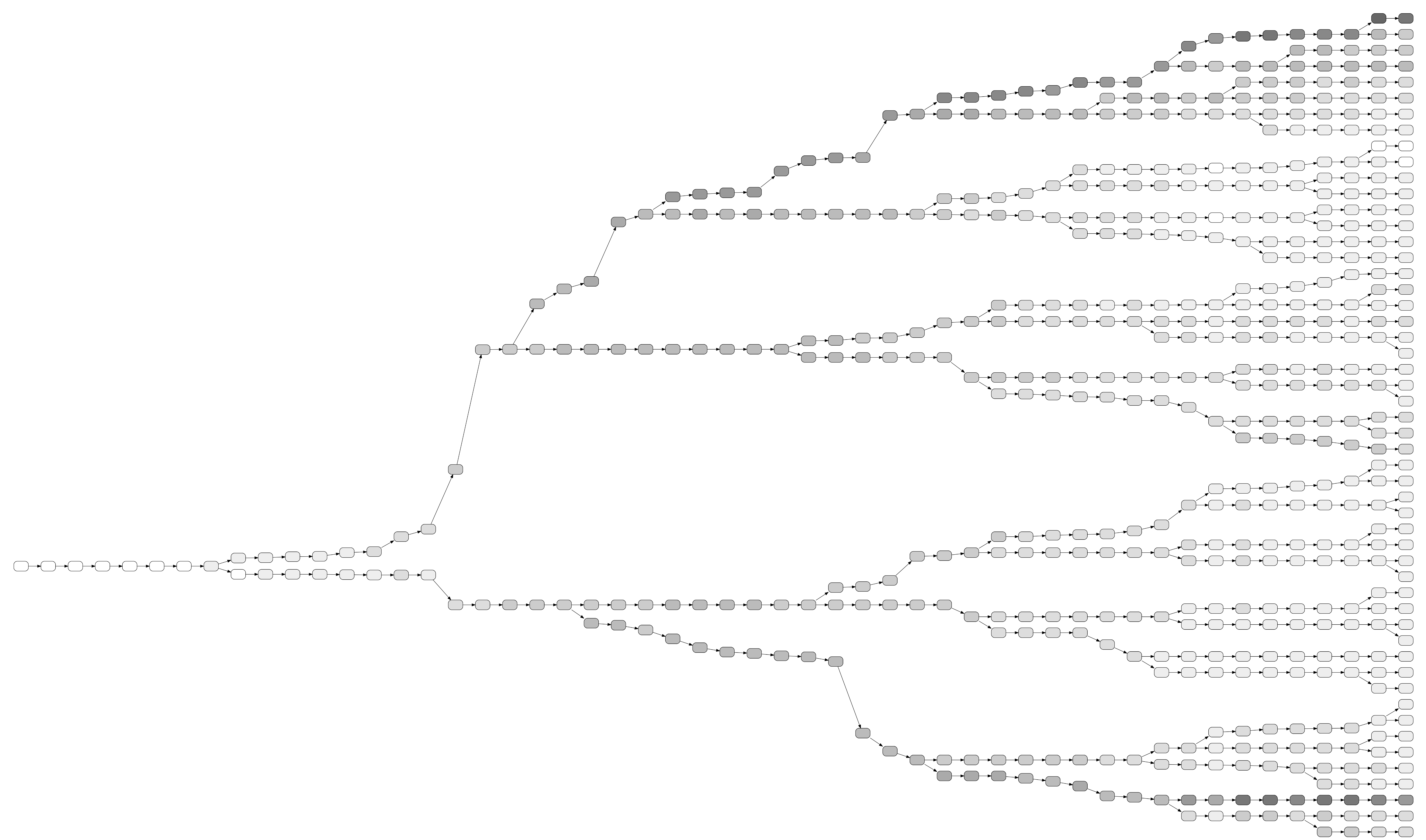}
\end{center}
\caption{Result of time-lapse microscopy and single cell analysis 
for the example of one cell lineage tree. 
Grey color indicates the fluorescence due to mCherry, which is correlated to the amount of plasmids in a cell. }\label{dataAnaTree}
\end{figure}

The employed data was derived from a time-lapse microscopy movie of growing {\it Bacillus megaterium} cells~\cite{manuskriptKarin}.
These cells were harboring a multi-copy plasmid that contained a xylose-inducible expression system~\cite{Stammen2010}.
The main components of this system consist of the xylose repressor gene \textit{xylR}, the operator region where the XylR protein can bind and a target gene.
In presence of xylose the repressor is removed which leads to an induction of expression of the target gene.
In the applied plasmid \textit{xylR} was fused with the \textit{mCherry} fluorescence gene thus when XylR-mCherry is bound the plasmid is tagged and can be visualized \textit{in vivo} via fluorescence microscopy.
At a sufficient plasmid copy number the signal is high enough to quantify the plasmid abundance in single cells.
Using image sequences it is possible to generate time-lapse movies and to follow up plasmid migration and segregation.
Spatial and temporal tracking of cells as well as quantification of fluorescence is done by image processing software \cite{Klein2012}.
The final result is a cell lineage tree (see Figure~\ref{dataAnaTree}).

We use these data to determine the mode of plasmid replication, and to determine
the plasmid transmission characteristics. We intend to 
validate the model structure 
and not to do a detailed data analysis; 
therefore, the parameter are estimated by rather naive methods.  
Based upon these
parameters, the model developed above predicts 
the plasmid distribution. The 
model predictions are compared with 
the experimental data.\par\medskip

\begin{figure}[htb]
\begin{center}
\includegraphics[width=13cm]{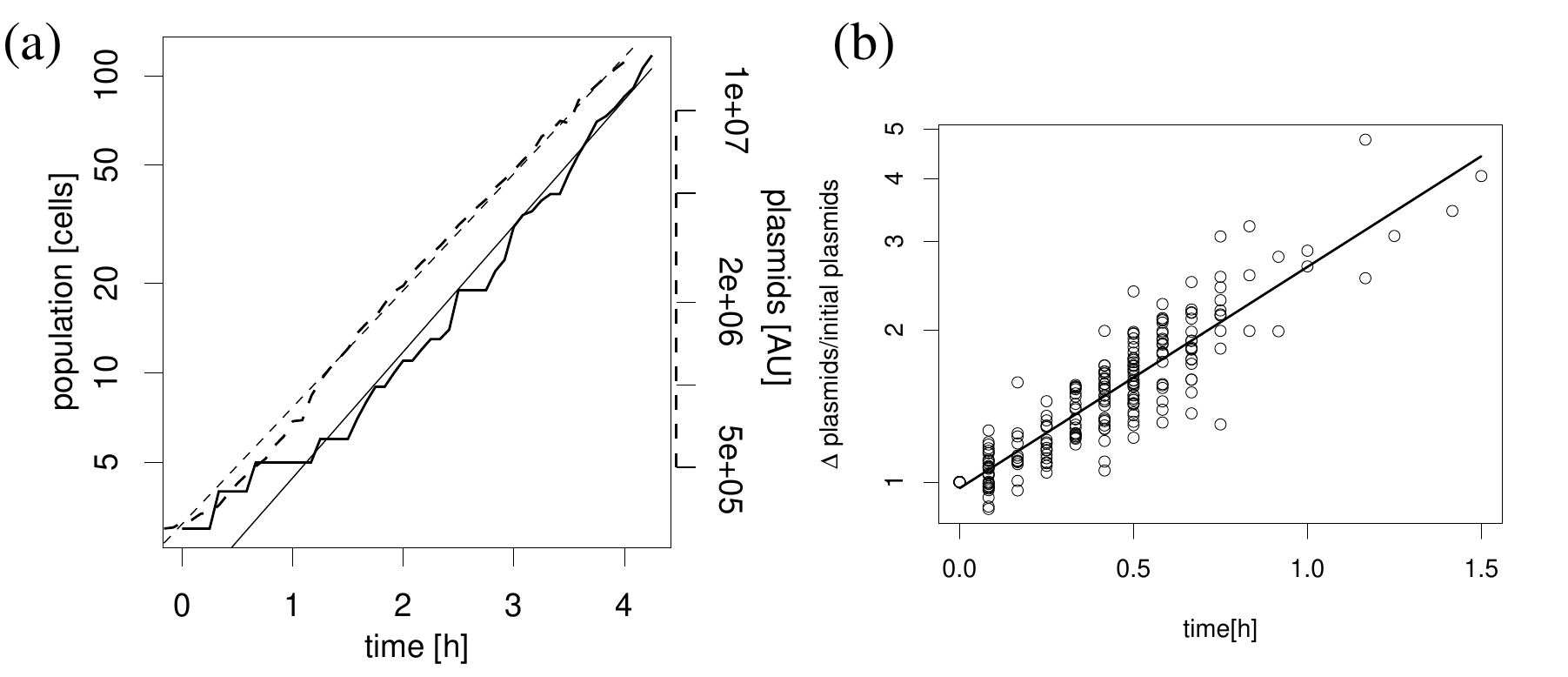}
\end{center}
\caption{(a) Population dynamics (solid line) and total plasmid content (summarized over all individuals, dashed line). Data: fat lines, exponential fit: thin lines. (b) Exponential fit 
of the relative plasmid increase between two cell division events.
}\label{dataAna}
\end{figure}

\paragraph{Cell and plasmid reproduction.} 
A first look at the time course of 
the bacterial and total plasmid population size
(we take the sum of all plasmids in all cells) show that plasmid population and cell population both grow exponentially with the
same exponent (Figure~\ref{dataAna}, left panel). This could be  
a first hint for cell controlled plasmid production  
 (recall the results of  section~\ref{simpleCase}). 
We use single cell data to investigate this idea. 
Surprisingly, a detailed analysis of the 
increase of plasmids during a cell cycle indicates an almost
perfect agreement
with an exponential grow  (Figure~\ref{dataAna}, right panel). Therefore 
we dismiss the hypothesis of cell controlled reproduction 
in favor of plasmid controlled reproduction. 
The semi-logarithmic, linear fit reveals for the population
growth the exponent $\beta = 0.975/h$, and the reproduction rate for plasmids $b_0=1.01/h$. The two exponents are almost identical, which is
an alternative explanation of the parallel increase of 
cells and plasmids in the left panel of figure~\ref{dataAna}. 
There is no obvious mechanism that couples plasmid- and cell reproduction, though it is rather unlikely that this precise agreement is pure coincidence. 
We take for the mortality $\mu=0$.

\begin{figure}[htb]
\begin{center}
\includegraphics[width=13cm]{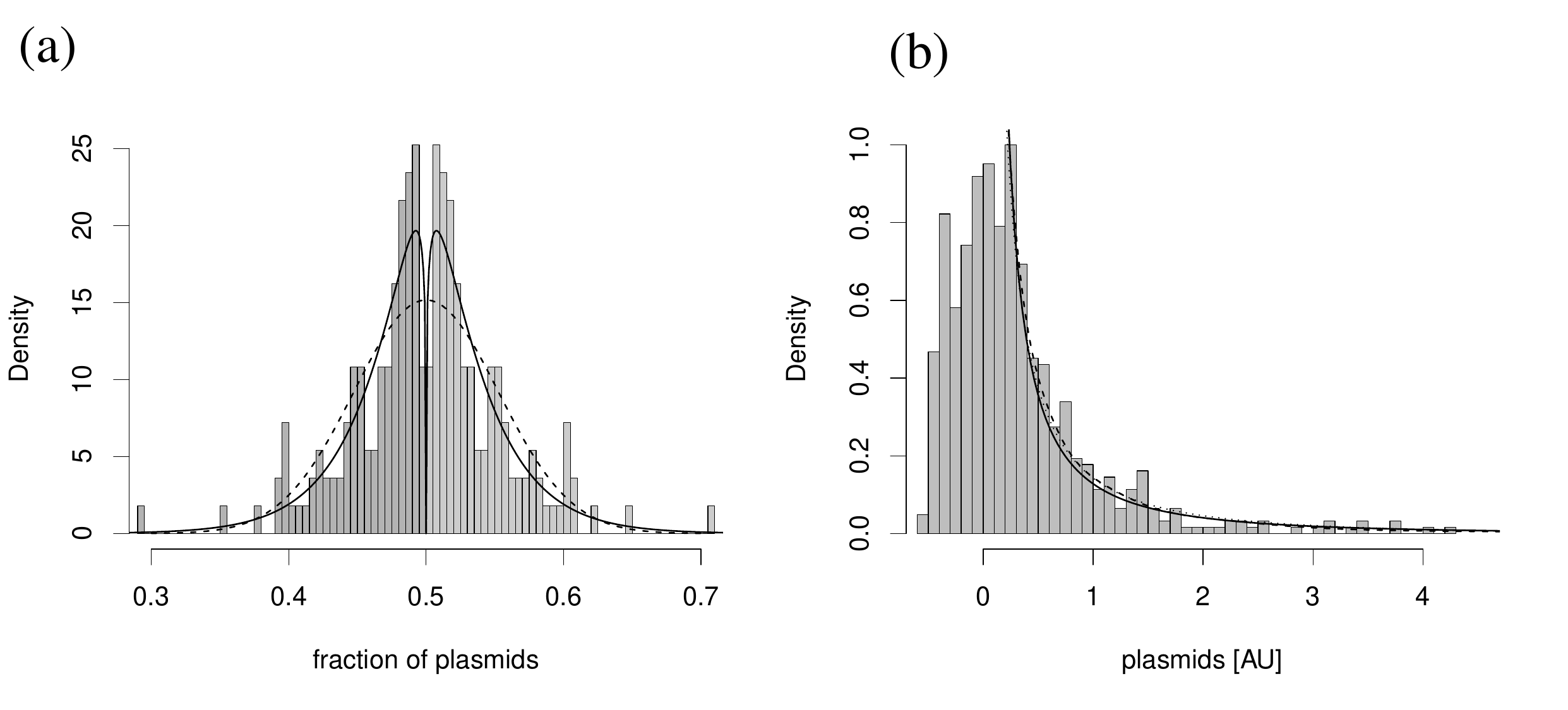}
\end{center}
\caption{
Left panel. Fraction of plasmids in the two daughter cells, together with a normal distribution 
(dashed line), 
and two gamma-distributions (solid lines), adapted to the data with more resp.\ with less than 50\% of  plasmids.
Right panel. Empirical distribution and model-based distribution 
for three different plasmid transmission kernels (solid: estimated kernel, dashed: symmetric $\delta$-peak, dotted: uniform kernel; note that the solid and the dashed line are almost on top of each other). 
}\label{simFig}
\end{figure}

\paragraph{Transmission characteristics of plasmids from mother to daughter.} 
As cell-tracking yields information about mother- and daughter cells, we
are able to compute the distribution of the relative fraction of plasmids in the two daughters, i.e.\ we
are able to produce a histogram for the kernel density 
$k(x,1)$. 
Figure~\ref{simFig} (a) displays this density, together
with a best fit of a normal distribution and a fit
of a gamma distribution for the cells with more
than $1/2$ of the mother's plasmid, resp.\ less
than $1/2$ of the mother's plasmids. The gap in 
the histogram at $1/2$ attracts attention. 
It is possible to interpret this gap as one indication for an unequal distribution of
plasmids between sister cells (see~\cite{manuskriptKarin} for a more detailed data analysis and discussion). The variance in 
the distribution is another indication for an unequal plasmid distribution between then two daughter cells. 
\par\medskip

\paragraph{Simulation of the data.} 
We feed these parameter in our population model. 
To compare the theoretical and the experimental distribution 
at the end of the experiment (after 4.5 h),
we shift the fluorescence distribution by 
a constant offset to the left, in order to compensate for auto-fluorescence (note that this shift is slightly inconsistent, as the plasmid reproduction rate is determined from the unshifted data); 
we furthermore rescale the theoretical plasmid distribution with a 
scalar factor in such a way that the 75\% quantile of 
 empirical
and simulated distribution agree
(see Figure~\ref{simFig}). 
The theoretical and empirical distributions seem to match nicely, 
in particular if we take into account that the data for low fluorescence are expected
to be rather noisy. 
Other transmission kernels as a $\delta$-peak or a uniform kernel
do not affect the outcome essentially. 
Our model seems to address the most fundamental principles of 
plasmid- and population dynamics in an appropriate way.\\
It is interesting to note that $b_0=\beta$ does not allow for 
an integrable equilibrium distribution. We expect a singularity
to appear at zero, and (as the average 
number of plasmids per cell is constant) at the same time few cells to increase infinitely the number of plasmids they inherit. 
This observation may correlate 
with experimental observations of many cells with few plasmids, 
and few cells that accumulate plasmids.

\section{Discussion}\label{discuss}

In this paper, we developed a model for plasmid dynamics 
in a bacterial population, based on ideas developed  in~\cite{bentley1993}. Using a continuum limit, 
we obtained the fragmentation equation, 
as e.g.\ proposed in~\cite{perthame:book}. 
The special
structure of our model allowed to convert 
the hyperbolic partial differential equation into an infinite set of ordinary differential equations for the moments. We then turned to investigate the 
shape of the equilibrium 
distributions of plasmids in dependence on different 
plasmid reproduction modes and plasmid transmission kernels. 
We defined two fundamental different plasmid 
reproduction modes: 
cell controlled production  
(a cell produces plasmids at a fairly constant 
rate, that is only decreased due to the plasmid load) 
and plasmid controlled reproduction (logistic
growth). Kuo et al.~\cite{kuo1996} name the latter mode 
mass-controlled production, and also 
introduce a third mode,
the ``division-controlled mode''. In the ``division-controlled mode'' cells duplicate plasmids during cell division, such that all daughter cells have -- from birth on-- the same amount of plasmids as the mother. 
From the dynamical point of view, this mode is 
less  interesting.\\ 
The analysis of the model indicates that the plasmid 
reproduction mode (production of plasmids per cell 
or reproduction of plasmids per plasmid) 
mainly influences  the shape of the distribution 
at few plasmids, 
while the plasmid reproduction velocity mainly 
influences  the distribution at the carrying capacity 
of a cell: We expect a pole of the distribution at 
zero plasmids in the plasmid controlled reproduction 
mode and that 
the distribution becomes small 
 at
small plasmid numbers for the cell controlled 
reproduction mode. At the carrying capacity, 
we expect the distribution to tend to zero if 
the plasmid reproduction rate  
is small in comparison with the cell reproduction rate, 
and to tend to infinity in the other case.
\par\medskip

Our results hint that the exact transmission 
mechanism of plasmids from mother to daughter is only
influential if it is distinctively unequal 
and plasmid reproduction is in the same range as cell 
reproduction. As the analysis of experimental data revealed, 
the latter requirement is given in biologically relevant 
systems. 
 In all other cases, the 
plasmid segregation mode only leads to a minor correction in the 
shape of the equilibrium distribution. This finding is 
an indication that the accumulation of plasmids observed in experiments is not solely due to unequal plasmid segregation,  
but also due to the interplay of plasmid 
reproduction and cell reproduction. We expect in particular that 
cells with a higher plasmid load will reproduce less fast and, in this way, plasmids may have a longer time to
accumulate within a cell. Therefore, these cells will 
divide even less often. 
In that, we find a positive feedback loop that offers
an second mechanism for accumulation, apart of 
unequal plasmid segregation. 

\noindent
{\bf Acknowledgements: }{\it We thank Lirike Neziraj for intensive discussions.
Part of this work was funded by the German Research Foundation (DFG)
within the priority program SPP1617 ``Phenotypic heterogeneity and sociobiology of bacterial populations''.}

\bibliographystyle{abbrv} 
\bibliography{contiLimitBib}

\end{document}